\documentclass[prd,twocolumn,nopacs,floatfix,amsmath,nofootinbib,amssymb,floatfix]{revtex4}
\usepackage{graphicx,color,dcolumn,booktabs,bm}
\usepackage{longtable,lscape}
\usepackage{pdfpages}
\usepackage{txfonts}
\usepackage{bbm}
\usepackage{overpic}
\usepackage{amssymb}
\usepackage{indentfirst}
\usepackage{feynmf}   
\usepackage{slashed}  
\usepackage{cases}
\usepackage{color}
\usepackage{multirow}
\usepackage{threeparttable}
\usepackage{epstopdf}
\usepackage{enumerate}
\usepackage[colorlinks, citecolor=blue,anchorcolor=red,menucolor=red, linkcolor=red,filecolor=red,urlcolor=blue,frenchlinks=red]{hyperref}

\graphicspath{{Figures/}} %

\begin{document}

\title{Investigating hybrid mesons with $0^{+-}$ and $2^{+-}$ exotic quantum numbers}

\author{Bing Chen$^{1,3}$}\email{chenbing@ahstu.edu.cn}
\author{Xiang Liu$^{2,3,4,5}$\footnote{Corresponding author}}\email{xiangliu@lzu.edu.cn}
\affiliation{ $^1$School of Electrical and Electronic Engineering, Anhui Science and Technology University, Bengbu 233000, China\\
$^2$School of Physical Science and Technology, Lanzhou University, Lanzhou 730000, China\\
$^3$Lanzhou Center for Theoretical Physics,
Key Laboratory of Theoretical Physics of Gansu Province,
Key Laboratory of Quantum Theory and Applications of MoE,
Gansu Provincial Research Center for Basic Disciplines of Quantum Physics, Lanzhou University, Lanzhou 730000, China\\
$^4$MoE Frontiers Science Center for Rare Isotopes, Lanzhou University, Lanzhou 730000, China\\
$^5$Research Center for Hadron and CSR Physics, Lanzhou University and Institute of Modern Physics of CAS, Lanzhou 730000, China}

\date{\today}

\begin{abstract}
Hybrid mesons occupy a unique position in the hadronic spectrum, offering a valuable insight into the non-perturbative nature of the glue field. Hybrid states with exotic $J^{PC}$ quantum numbers, such as the $1^{-+}$, $0^{+-}$ and $2^{+-}$, cannot mix with conventional $q\bar{q}$ mesons, making them critical for establishing the full spectrum of hybrid mesons. In this work, we investigate the masses and the strong decays of $0^{+-}$ and $2^{+-}$ hybrid mesons, which could be regarded as the first orbitally excited states of the $\pi_1(1600)$ and $\eta_1(1855)$ mesons. We develop a constituent gluon model and apply it to study the mass spectra and decay modes of these exotic states. Our results suggest that experiment may searched for these $0^{+-}$ and $2^{+-}$ hybrid mesons in the mass region between 2.0 GeV and 2.4 GeV. Additionally, we identify some key decay channels which should be targeted in future experiments.
\end{abstract}



 \maketitle

\section{Introduction}\label{sec1}

The non-perturbative behavior of strong interaction at low energies remains incompletely understood, presenting a significant challenge in modern particle physics \cite{Gross:2022hyw}. In Quantum chromodynamics (QCD), gluons, like quarks, carry color charge and directly experience color forces. This property allows the valence gluons to be the degrees of freedom within the hadronic systems, such as the glueball and hybrid states. The mutual attraction between gluons is believed to be intimately connected to the confinement property of QCD \cite{Close:1987er}, making the study of glueballs and hybrids crucial for understanding the behavior of soft gluonic fields. According to the lattice QCD calculations \cite{Chen:2005mg,Athenodorou:2020ani}, the lightest glueballs are expected to be the scalar ($0^{++}$) and tensor ($2^{++}$) states. However, these states may mix with the isoscalar $q\bar{q}$ mesons, which complicates their identification. In contrast, the low-lying hybrid mesons with exotic quantum numbers, such as $1^{-+}$, $0^{+-}$, and $2^{+-}$, cannot mix with conventional $q\bar{q}$ mesons. Studying them would be a key step toward establishing the complete spectrum of hybrid states.

A significant progress has been made in search for the hybrid mesons in recent years. The long-standing puzzle of the $\pi_1(1400)$ and $\pi_1(1600)$ states has been resolved through coupled-channel analyses, which indicate that only the $\pi_1(1600)$ is necessary to describe experimental data \cite{JPAC:2018zyd,Kopf:2020yoa}. In 2023, the BESIII Collaboration discovered a new $1^{-+}$ state \cite{BESIII:2022iwi,BESIII:2022riz}, the $\eta_1(1855)$, which could be the isospin partner of the $\pi_1(1600)$. Recently, BESIII tried to search for the $\eta_1(1855)$ state in the $\chi_{c1}\to\eta\eta_1(1855)\to\eta\eta\eta^\prime$ \cite{BESIII:2025izw}. But no significant $\eta_1(1855)$ was observed. These experimental results offer the valuable benchmarks for constructing the phenomenological models of hybrid mesons.

On the theoretical side, hybrid mesons have also attracted considerable attention in the past years. The mass spectrum and strong decays of low-lying $c\bar{c}g$ states have been explored using a constituent gluon model \cite{Farina:2020slb}. The flavor mixing of light hybrid mesons has been investigated in Ref. \cite{Swanson:2023zlm}, where the possible mixing between hybrid and conventional mesons was also discussed. Lattice QCD has been employed to study the light $1^{-+}$ hybrid mesons \cite{Woss:2020ayi,Liang:2024lon,Chen:2022isv} and the strong decays of the $\eta_{c1}(1^{-+})$ meson \cite{Shi:2023sdy}. Lattice QCD was also taken for analyzing the hybrid spin-dependent and hybrid-quarkonium mixing potentials appearing at order $(1/m_Q)^1$ \cite{Schlosser:2025tca}, where the gluon spin was considered as  $\kappa^{PC}=1^{+-}$ for a hybrid meson. The QCD sum rule method has been applied to study the masses and strong decays of light hybrid mesons, particularly those with exotic quantum numbers \cite{Wang:2023whb,Tan:2024grd,Barsbay:2024vjt,Esmer:2025xss}. Recently, the QCD sum rule was also used to investigate the masses and strong decays of $c\bar{c}g$ and $b\bar{b}g$ states \cite{Wang:2024hvp,Barsbay:2025vjq}. Within the Born-Oppenheimer effective field theory, the properties of heavy $Q\bar{Q}g$ states have been widely studied, including their masses, decays, hyperfine splittings, and mixing with conventional $Q\bar{Q}$ states (see \cite{Berwein:2024ztx} and references therein). Following the observation of the $\eta_1^\prime(1855)$, numerous theoretical works have explored the possibility that the $\pi_1(1600)$ and $\eta_1^\prime(1855)$ states belong to the SU$(3)$ flavor $1^{-+}$ multiplet \cite{Qiu:2022ktc,Shastry:2022mhk,Chen:2022qpd,Chen:2023ukh}. These developments underscore the growing interest in hybrid mesons among the theorists.

As mentioned above, the hybrid mesons with exotic quantum numbers are particularly important for establishing the complete spectrum of hybrid states. The $\pi_1(1600)$ and $\eta_1^\prime(1855)$ are the strong candidates for the SU$(3)$ flavor $1^{-+}$ multiplet, serving as the reference states for further exploration of the undiscovered hybrids. In fact, the lattice QCD calculations indicated that the light hybrid mesons with $J^{PC}=(0,~2)^{+-}$ are about $300\sim400$ MeV above the $1^{-+}$ states  \cite{Dudek:2011bn,Dudek:2013yja}. Up to now,  several studies have focused on the properties of these light hybrid mesons with exotic quantum numbers. For instance, the strong decays of light $q\bar{q}g$ states with $J^{PC}=0^{+-}$ and $J^{PC}=2^{+-}$ have been investigated using the flux tube model \cite{Close:1994hc}. Combing with the heavy quark expansion of QCD, Page, Swanson, and Szczepaniak developed the flux tube model \cite{Page:1998gz}. They applied the model to study the strong decays of low-lying hybrid mesons, including the $1^{-+}$, $0^{+-}$ and $2^{+-}$ states. However, the flux tube model predicts only one $2^{+-}$ state, which is inconsistent with the lattice QCD results \cite{Dudek:2011bn,HadronSpectrum:2012gic}.

A constituent gluon model which was proposed by Swanson and Szczepaniak \cite{Swanson:1998kx} offers a more realistic description of hybrid meson systems. Recently, the constituent gluon model was refined and applied for studying the charmonium hybrid \cite{Farina:2020slb}. In this model, an axial constituent gluon ($J^{P_gC_g}=1^{+-}$) was used to represent low-lying gluonic degrees of freedom. This picture is partially supported by lattice QCD computations \cite{Dudek:2011bn}.
In this work, we further develop the constituent gluon model and employ it to study the masses and strong decays of light $0^{+-}$ and $2^{+-}$ hybrid states.
We begin by testing the model with a simple harmonic oscillator (SHO) potential in Sec. \ref{sec2}, demonstrating its ability to reproduce the mass gaps of low-lying gluelump states. We improve the model by employing the more reliable potential for the hybrid state. By combing with the results of lattice QCD, we further predict the masses of $0^{+-}$ and $2^{+-}$ states. The strong decays of these states are discussed in Sec. \ref{sec3}, and the paper concludes with a summary and outlook in Sec. \ref{sec4}. Some detailed calculations are collected in the appendix.

\section{Mass spectrum}\label{sec2}

\subsection{The constituent gluon model with SHO potential}\label{sec21}

The quark potential model has achieved a remarkable success in describing the mass spectrum of ordinary hadrons, including $q\bar{q}$ mesons \cite{Godfrey:1985xj} and $qqq$ baryons \cite{Capstick:1986ter}.\footnote{One could consult Refs. \cite{Mukherjee:1993hb,Lucha:1991vn} and the papers citing these reviews for more details of the quark potential model.} Inspired by the success of quark model, a key topic in the study of hybrid mesons is identifying the adiabatic potentials which can effectively describe the interactions between a quark-antiquark pair in the presence of gluonic excitations.

The adiabatic potential has been extensively studied using the lattice simulations over the years. Griffiths \textit{et al.} first discussed the potentials which correspond to the lowest two adiabatic gluonic surfaces \cite{Griffiths:1983ah,Campbell:1984fe}. With advancements in lattice methods, the static quark potentials for more gluonic configurations were obtained \cite{Juge:1997nc,Juge:2002br}. Recently, more precise calculations of hybrid static potentials were performed in SU(3) lattice gauge theory \cite{Capitani:2018rox,Schlosser:2021wnr,Alasiri:2024nue}. These lattice QCD results provide valuable insights for constructing phenomenological models of hybrid mesons.

\begin{figure}[htbp]
\centering
\includegraphics[width=8.5cm,keepaspectratio]{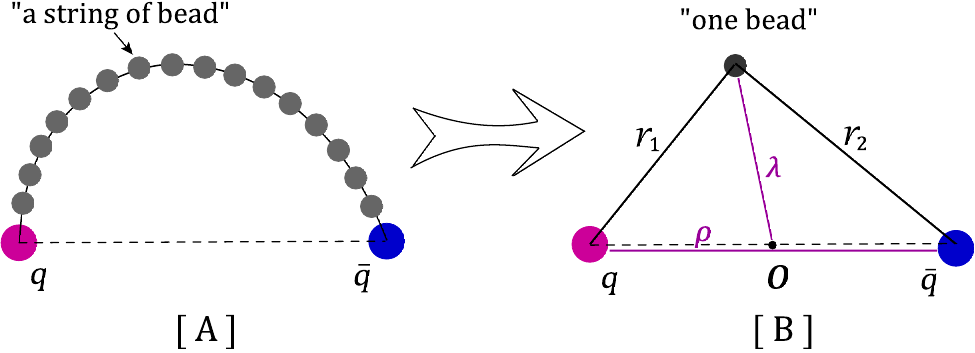}
\caption{Schematic diagram of a hybrid meson in the flux tube model. $[A]$: The confining interaction between the quark-antiquark pair is depicted as a series of mass beads. $[B]$: A simplified representation where the confining potential is modeled by a single bead captures the essential features of the interaction.}
\label{fluxtube}
\end{figure}

On the other hand, several kinds of models have attempted to reproduce these static potentials, including the flux tube model \cite{Isgur:1984bm}, the QCD string model \cite{Kalashnikova:2002tg}, and the constituent gluon model \cite{Swanson:1998kx}. Among these models, the flux tube model is particularly notable for its ability to describe not only the ordinary $q\bar{q}$ and $qqq$ hadrons but also the exotic states such as the hybrids, glueballs, and multiquark states \cite{Isgur:1984bm}. The flux tube model is based on the strong-coupling Hamiltonian of lattice QCD, where gluonic degrees of freedom are transformed into collective string-like flux tubes (see Fig. \ref{fluxtube}). A small oscillation approximation was employed to derive relatively simple analytical results \cite{Isgur:1984bm}. However, this approximation may be inadequate for light hadrons, as it overestimates the potential gap between the lowest excited and ground gluonic fields (see Fig. \textcolor{red}{2} of Ref. \cite{Barnes:1995hc}).

To address this limitation, Barnes \textit{et al.} refined the flux tube model by simplifying the string to a single bead \cite{Barnes:1995hc}. Their improved model accurately reproduced the spectrum of conventional quarkonia. However, it predicted a larger mass difference between the first orbitally excited and ground $c\bar{c}g$ multiplets by comparing the lattice QCD results \cite{Capitani:2018rox}.\footnote{A smaller orbital excitation gap is expected for hybrid systems due to their relatively flat adiabatic potentials.}

The hybrid meson is simplified to be a three-body system when the string of mass beads is replaced by a single bead. This picture aligns with the QCD string model \cite{Kalashnikova:2002tg}\footnote{The hybrid adiabatic potentials derived from the QCD string model show good agreement with the lattice results \cite{Kalashnikova:2002tg}.} and the constituent gluon model \cite{Swanson:1998kx}. Based on these models, we could develop a three-body hybrid model in which the massive bead in the flux tube model (see $[B]$ diagram in Fig. \ref{fluxtube}) is replaced by a transverse electric (TE) gluon ($J_g^{P_gC_g}=1^{+-}$).\footnote{This picture is partially supported by lattice QCD studies since the gluonic structure in the operators of low-lying hybrid states overlaps well with the chromomagnetic component \cite{Dudek:2011bn}. The bag model also suggests that the lowest excited gluonic field in a hybrid meson should be a $1^{+-}$ state \cite{Jaffe:1975fd,Barnes:1982tx}.}

\begin{figure}[htbp]
\centering
\includegraphics[width=4.6cm,keepaspectratio]{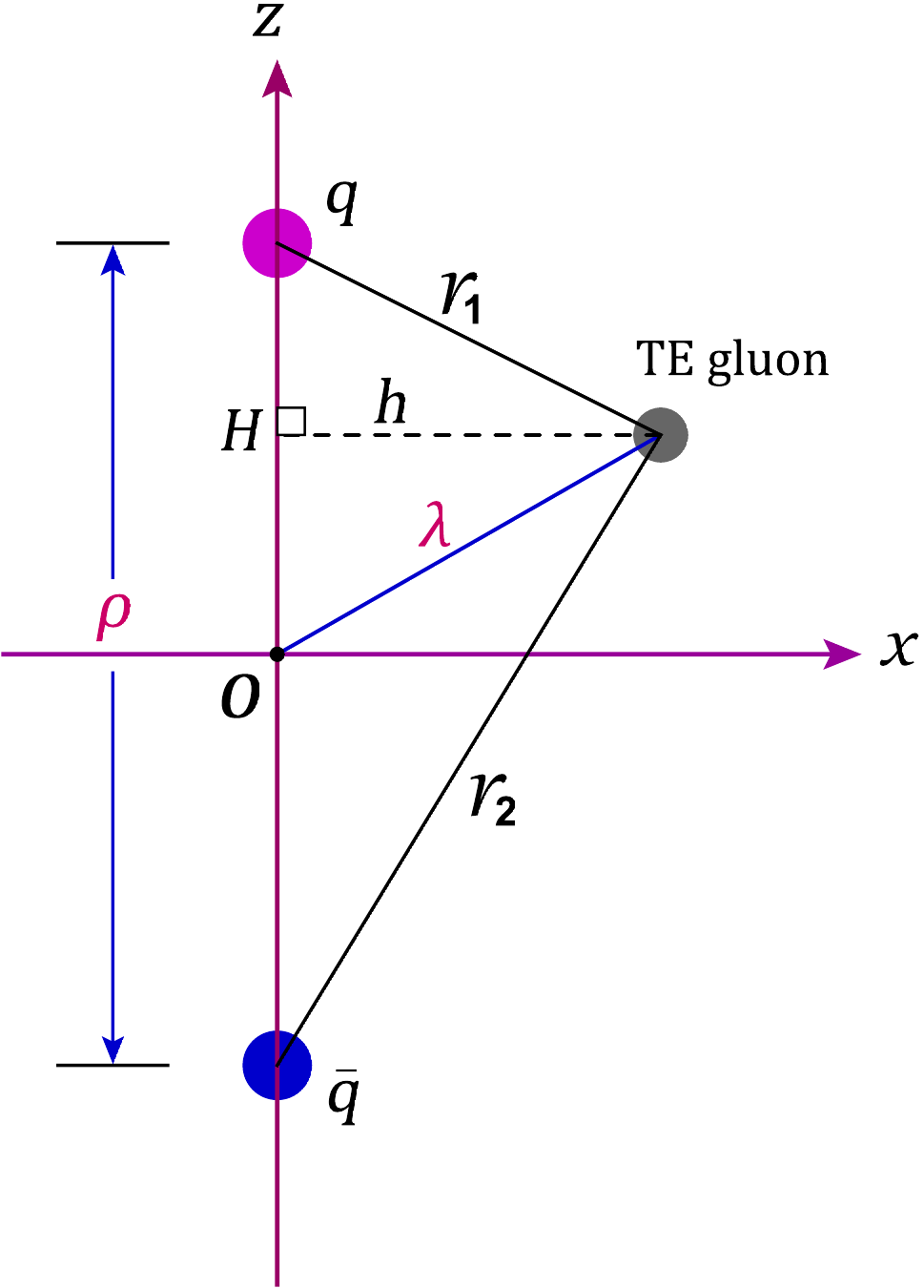}
\caption{A low-lying hybrid meson in the constituent gluon model. The lowest excited gluonic field in the hybrid system is approximated as a TE gluon.}
\label{hybrid}
\end{figure}

As shown in Fig. \ref{hybrid}, the long-range confinement potential is assumed to exist only between the TE gluon and the quark. Then the Schr\"{o}dinger equation for a hybrid state is given by
\begin{equation}
\left(\sum\limits_{i=q, \bar{q}, g} \frac{\textbf{p}_i^2}{2m_i} + \sum\limits_{j=1}^{2}V_\textup{eff.}(r_{j}) \right)\psi_H = E_H \psi_H.  \label{eq1}
\end{equation}
To test the constituent gluon model, we first employ a simple harmonic oscillator (SHO) potential as the confinement interaction. The effective potential $V_\textup{eff.}(r_{j})$ is written as
\begin{equation}
\sum\limits_{j=1}^{2}V_\textup{eff.}(r_{j}) = \textsl{C}_{jg}\left( \frac{1}{2}Kr_1^2 + \frac{1}{2}Kr_2^2 \right) + c_H. \label{eq2}
\end{equation}
Here, $K$ is the spring constant representing the strength of the confinement interaction. $\textsl{C}_{jg} = 3/2$ is the color coefficient for hybrid mesons, and $c_H$ is a mass-renormalization constant. Using the relationships
\begin{equation}
r_1^2 = \left( \frac{\rho}{2} - z \right)^2 + h^2, \quad r_2^2 = \left( \frac{\rho}{2} + z \right)^2 + h^2, \nonumber
\end{equation}
the effective potential $V_\textup{eff.}(r_{j})$ can be reformulated as
\begin{equation}
V_\textup{eff.}(\rho,\lambda) = \textsl{C}_{ig}\left( \frac{1}{4}K\rho^2 + K\lambda^2 \right) + c_H. \label{eq3}
\end{equation}
Here, Jacobi coordinates are used to simplify the three-body problem. The Schr\"{o}dinger equation then becomes
\begin{equation}
\left[ \frac{\textbf{p}_\rho^2}{2m_\rho} + \frac{\textbf{p}_\lambda^2}{2m_\lambda} +\frac{3}{2}K\left( \frac{\rho^2}{4} + \lambda^2 + c_M \right) \right]\psi_H(\rho,\lambda) = E_H \psi_H(\rho,\lambda),  \label{eq4}
\end{equation}
where the reduced masses are defined as
\begin{equation}
m_\rho = \frac{m_1m_2}{m_1+m_2}, \quad m_\lambda = \frac{(m_1+m_2)m_g}{m_1+m_2+m_g}. \nonumber
\end{equation}
As shown in Eq. (\ref{eq4}), the $\rho$-mode and $\lambda$-mode are independent when the SHO potential is used for the long-range confinement interaction between the TE gluon and the quark.

Interestingly, the hybrid wave function is often approximated as a product of the quark and gluon wave functions \cite{Farina:2020slb}. This ansatz is supported by Eq. (\ref{eq4}). Specifically, the wave function of a hybrid state can be written as
\begin{equation}
\Psi_{n_{q\bar{q}}L_{q\bar{q}},n_gL_g}(\bm{\rho},\bm{\lambda}) = R_{n_{q\bar{q}}L_{q\bar{q}}}(\rho)Y_{L_{q\bar{q}}m_{q\bar{q}}}(\hat{\bm{\rho}})R_{n_gL_g}(\lambda)Y_{L_gl_g}(\hat{\bm{\lambda}}).  \label{eq5}
\end{equation}
Here, $R_{nl}(r)Y_{lm}(\hat{\bm{r}})$ is the spatial wave function of the SHO potential. This wave function was also used in our recent work \cite{Chen:2023ukh} to study the decay behaviors of $\pi_1(1600)$ and $\eta_1(1855)$. In the next subsection, we will use the SHO wave function as a trial function to solve the Schr\"{o}dinger equation with a more realistic potential, such as the linear confinement potential.

Since the $\rho$-mode and $\lambda$-mode of a hybrid meson are independent in the SHO model, the quark-antiquark pair provides a static color octet source for the gluonic field at their center. Thus, we may test this simple hybrid model by comparing the $\lambda$-mode excitation energy with gluelump masses obtained from other methods.

The energy level of a hybrid meson in the SHO potential is given by
\begin{equation}
E_H = \left( 2n_\lambda +l_\lambda + \frac{3}{2} \right)\omega_\lambda +\left( 2n_\rho +l_\rho + \frac{3}{2} \right)\omega_\rho.  \label{eq6}
\end{equation}
The angular frequencies $\omega_\lambda$ and $\omega_\rho$ are defined as
\begin{equation}
\omega_\lambda = \sqrt{\frac{3K}{m_\lambda}}, \quad \omega_\rho = \sqrt{\frac{3K}{4m_\rho}}. \nonumber
\end{equation}
For comparison, we also express the energy level of light mesons using the SHO potential. The Schr\"{o}dinger equation for conventional $q\bar{q}$ mesons is
\begin{equation}
\left[ \frac{\textbf{p}^2}{2m_\mu} + \textsl{C}_{q\bar{q}}\left( \frac{1}{2}Kr^2 \right) + C_M \right]\psi_M(r) = E_M \psi_M(r),  \label{eq7}
\end{equation}
where $\textsl{C}_{q\bar{q}} = 4/3$ is the color coefficient for mesons. The energy level of a light meson in the SHO potential is
\begin{equation}
E_M = \left( 2n + l + \frac{3}{2} \right)\omega_M,  \label{eq8}
\end{equation}
with $\omega_M = \sqrt{\frac{4K}{3m_\mu}}$. Comparing $\omega_\lambda$ and $\omega_M$, we obtain
\begin{equation}
\omega_\lambda = \frac{3}{4}\omega_M\sqrt{1+2\frac{m_q}{m_g}}.  \label{eq9}
\end{equation}
Using the masses of $\pi$, $\rho(770)$, $h_1(1170)$, and $b_1(1235)$ \cite{ParticleDataGroup:2024cfk}, we estimate $\omega_M$ as
\begin{equation}
\omega_M \simeq \frac{M(h_1)+M(b_1)}{2}-\frac{M(\pi)+3M(\rho)}{4} \approx 0.582~\textup{GeV}. \nonumber
\end{equation}
The $\pi_1(1600)$ is approximately 1.0 GeV above the average mass of $\pi$ and $\rho$ states. Similarly, the lowest heavy hybrid states are predicted to be 1.0 GeV above the $1S$ $Q\bar{Q}$ mesons \cite{Barnes:1995hc}. Lattice QCD results \cite{Capitani:2018rox,HadronSpectrum:2012gic,Ryan:2020iog} and the constituent gluon model \cite{Farina:2020slb} suggest that the lowest $Q\bar{Q}g$ hybrids are about 1.3 GeV above the $1S$ $Q\bar{Q}$ mesons. Based on these results, we take $m_g = 1.05$ GeV and $m_q = 0.32$ GeV, yielding
\begin{equation}
\omega_\lambda = \frac{3}{4}\times 0.582 \sqrt{1+2\times\frac{0.32}{1.05}} = 0.554~\textup{GeV}.  \label{eq10}
\end{equation}

From a phenomenological perspective, the $2^{--}$ and $3^{+-}$ gluelumps can be viewed as the orbital excitations of $1^{+-}$ state \cite{Karl:1999wq,Kalashnikova:2002tg}. Various methods have predicted the mass splittings $\delta_{21} = M_{2^{--}} - M_{1^{+-}}$ and $\delta_{31} = M_{3^{+-}} - M_{1^{+-}}$. Table \ref{table1} summarizes results from the bag model \cite{Karl:1999wq}, the lattice QCD \cite{Foster:1998wu}, the QCD string model \cite{Simonov:2000ky}, the mean-field approach in Coulomb gauge QCD \cite{Guo:2007sm}, and the transverse constituent gluon model \cite{Buisseret:2008pd}. Most predictions for $\delta_{21}$ are slightly above 0.50 GeV, consistent with the value of $\omega_\lambda$ in Eq. (\ref{eq10}). Additionally, the $\delta_{31}$ values are approximately twice the corresponding $\delta_{21}$, aligning with the SHO model.

The conclusion above would be solid even the uncertainties of $m_q$ and $m_g$ are considered. Specifically, alteration of $\omega_\lambda$ not exceeds 30 MeV, when $m_q$ changes from $0.28$ GeV to $ 0.38$ GeV and $m_g$ changes from $0.85$ GeV to $ 1.30$ GeV. In fact, the value of $\omega_\lambda$ in Eq. (\ref{eq10}) might be slightly overestimated due to the anomalously light mass of pion meson. In the next subsection, we will replace the SHO potential with a linear confinement potential to construct a more realistic model for hybrid meson masses.

\begin{table}[htbp]
\caption{Predicted mass gaps of the $2^{--}/3^{+-}$ and $1^{+-}$ gluelumps from Refs. \cite{Karl:1999wq,Foster:1998wu,Simonov:2000ky,Guo:2007sm,Buisseret:2008pd} (GeV).
}\label{table1}
\renewcommand\arraystretch{1.20}
\begin{tabular*}{86mm}{c@{\extracolsep{\fill}}ccccc}
\toprule[1pt]\toprule[1pt]
\multirow{2}{*}{$J^{PC}$}   & \multicolumn{5}{c}{$M(J^{PC})-M(1^{+-})$~~[GeV]}     \\
\cline{2-6}
                  &    Ref.~\cite{Karl:1999wq}  &  Ref.~\cite{Foster:1998wu}   &    Ref.~\cite{Simonov:2000ky}  &   Ref.~\cite{Guo:2007sm} &   Ref.~\cite{Buisseret:2008pd}  \\
\toprule[1pt]
  $2^{--}$        &    0.54                     &   0.567       &   0.49       & 0.59    &   0.57        \\
  $3^{+-}$        &    1.01                     &   0.972       &   0.84       & 1.11    &   0.97        \\
\bottomrule[1pt]\bottomrule[1pt]
\end{tabular*}
\end{table}

\subsection{The constituent gluon model with Cornell potential}\label{sec22}

The Cornell potential has been widely used to describe the spectrum of conventional hadrons. It consists of two parts: a Coulomb potential at short distances and a linear confinement potential at long distances \cite{Eichten:1978tg}. For hybrid mesons, we replace the SHO potential in Eq. (\ref{eq2}) by the linear confinement potential and include a Coulomb potential for the quark-antiquark pair due to one-gluon exchange. Thus, the effective interaction potential for the hybrid meson is given by
\begin{equation}
\begin{aligned}\label{eq11}
V_{q\bar{q}g}(r_1,r_2,\rho) =~& \textsl{C}_{jg}\left( \frac{3}{4}br_1 + \frac{3}{4}br_2 \right) + \textsl{C}^\prime_{q\bar{q}}\frac{\alpha_s}{\rho} + c^\prime_H \\
                                                 =~& \frac{9}{8}b(r_1 + r_2 )+\frac{1}{6}\frac{\alpha_s}{\rho} + c^\prime_H.
\end{aligned}
\end{equation}
Here, $\textsl{C}^\prime_{q\bar{q}} = 1/6$ is a color coefficient for a hybrid meson. The Coulomb term in Eq. (\ref{eq11}) ensures the repulsive behavior of the hybrid static potential at short distances between the quark and antiquark, as confirmed by lattice QCD \cite{Schlosser:2021wnr}. We now solve the following Schr\"{o}dinger equation:
\begin{equation}
\left[ \frac{\textbf{p}_\rho^2}{2m_\rho} + \frac{\textbf{p}_\lambda^2}{2m_\lambda} + \frac{9}{8}b(r_1 + r_2 )+\frac{1}{6}\frac{\alpha_s}{\rho} + c^\prime_H \right]\psi_H = E_H \psi_H.  \label{eq12}
\end{equation}
We neglect the spin-dependent interactions since even the ground hybrid mesons have not been firmly established.\footnote{According to lattice QCD results \cite{Dudek:2011bn}, the light-flavor hybrid meson supermultiplet includes $0^{-+}$, $1^{-+}$, $1^{--}$, and $2^{-+}$ states. Currently, only the isovector $\pi_1(1600)$ and isoscalar $\eta_1(1855)$ with $J^{PC}=1^{-+}$ have been experimentally confirmed.} Therefore, the results from Eq. (\ref{eq12}) can be interpreted as the spin-averaged masses of hybrid multiplets.

As mentioned earlier, we use a simple variational method to solve Eq. (\ref{eq12}), with the SHO wave function (Eq. (\ref{eq5})) as the trial wave function. This approach has been shown to yield accurate results for solving the Schr\"{o}dinger equation of two- and three-body systems \cite{Karl:1994ji,Schoberl:1984ha}.\footnote{Compared to the result of the Gaussian expansion method, the prediction of variational method is overestimated by several MeV.} For the ground hybrid states \cite{Eichten:1978tg}, the trial wave function is written as
\begin{equation}
\Psi_{00,00}(\bm{\rho},\bm{\lambda}) = \frac{\beta_\rho^{3/2}\beta_\lambda^{3/2}}{\pi^{3/2}} \textup{e}^{ -\left( \beta_\rho^2 \rho^2 + \beta_\lambda^2 \lambda^2 \right)/2 }.  \label{eq13}
\end{equation}
The energy of the ground states is given by
\begin{equation}
E_{00,00} = \frac{3}{4} \left( \frac{\beta_\rho^2}{m_\rho} + \frac{\beta_\lambda^2}{m_\lambda} \right) + \frac{9b}{8\sqrt{\pi}} \sqrt{\frac{1}{\beta_\rho^2} + \frac{4}{\beta_\lambda^2}} + \frac{\alpha_s \beta_\rho}{3\sqrt{\pi}} + c^\prime_H.  \label{eq14}
\end{equation}
The detailed calculations are provided in Appendix \ref{A}, along with the energy of the $1\rho$ excited states.

Since the hybrid mesons is not yet fully established, we cannot directly fix the parameters using the measured masses of known hybrids. Instead, we assume that the parameters for light hybrid mesons are the same as those for conventional $q\bar{q}$ mesons. We validate this assumption by comparing our results with lattice QCD predictions and experimental evidence. To this end, we express the potential for $q\bar{q}$ mesons as
\begin{equation}
V_{q\bar{q}}(r) =  C_{q\bar{q}}\left( -\frac{\alpha_s}{r} + \frac{3}{4}br \right) + c_M.   \label{eq15}
\end{equation}
This is the well-known Cornell potential, widely used to study meson masses. In our previous work \cite{Chen:2023ukh}, the parameters $\alpha_s$ and $b$ were fixed as
\begin{equation}
\alpha_s = 0.640, \quad b = 0.165~\textup{GeV}, \nonumber
\end{equation}
with $m_{u,d} = 0.320$ GeV and $m_s = 0.450$ GeV. For hybrid mesons, we take $c^\prime_H = 1.890$ GeV and $1.820$ GeV for the $q\bar{q}g$ and $s\bar{s}g$ systems, respectively. The predicted masses for the low-lying light hybrid mesons are listed in Table \ref{table2}.

\begin{table}[htbp]
\caption{Predicted average masses of the low-lying $q\bar{q}g$ and $s\bar{s}g$ multiplets (MeV).} \label{table2}
\renewcommand\arraystretch{1.2}
\begin{tabular*}{86mm}{@{\extracolsep{\fill}}ccccc}
\toprule[1pt]\toprule[1pt]
 $(n_\lambda l_\lambda,~n_\rho l_\rho)$    &  $(00,~00)$   &  $(00,~01)$  &  $(01,~00)$  &   $(00,~10)$    \\
\toprule[1pt]
  $\bar{M}_{q\bar{q}g}$                    &  1755         &   2118       & 2226         &   2113 \\
  $\bar{M}_{s\bar{s}g}$                    &  1949         &   2266       & 2394         &   2255 \\
\bottomrule[1pt]\bottomrule[1pt]
\end{tabular*}
\end{table}

The average mass of the ground $q\bar{q}g$ states is determined to be 1775 MeV, which is approximately 100--150 MeV higher than the mass of the $\pi_1(1600)$. This result is consistent with predictions from the constituent gluon model \cite{Kalashnikova:1993xb} and lattice QCD \cite{Dudek:2011bn}. Additionally, the average mass of the ground $q\bar{q}g$ states is about 200 MeV below that of the $s\bar{s}g$ states, which aligns with the mass difference between the $\pi_1(1600)$ and the $\eta_1(1855)$. Here, a small SU(3) flavor violation is expected to exist in a light flavor hybrid nonet. This SU(3) flavor violation is well-known in the meson and baryon sectors. For example, the mass of $\phi(1020)$ is about 130 MeV heavier than than the $K^\ast(892)$. In the constituent quark model, the mass gap of $\phi(1020)$ and $K^\ast(892)$ states could be mainly attributed to the mass difference of up/down and strange quarks.

We focus on the masses of $q\bar{q}g$ and $s\bar{s}g$ states with exotic quantum numbers $0^{+-}$ and $2^{+-}$. In our framework, these states belong to the $(n_\lambda l_\lambda,~n_\rho l_\rho) = (00,~01)$ multiplet. As shown in Table \ref{table2}, the average mass of the $(n_\lambda l_\lambda,~n_\rho l_\rho) = (00,~01)$ multiplet is approximately 360 MeV above the ground states. Interestingly, a signal of an $h_{2l}$ state around 2.0 GeV was potentially observed at the multi-particle spectrometer facility at Brookhaven National Laboratory (BNL) in 2005 \cite{Adams:2005tx}. Specifically, a resonance structure with exotic quantum numbers $I^GJ^{PC} = 0^-2^{+-}$ was identified in the process $\pi^- + p \to h_2(0^-2^{+-}) + n \to b_1\pi n \to \omega\pi\pi n \to 5\pi n$. According to the E852 experiment, the $h_{2l}$ state is about 350 MeV above the $\pi_1(1600)$, which is consistent with our predictions. Furthermore, the central mass of the first radially excited $q\bar{q}g$ state is predicted to be 2113 MeV. This result supports the interpretation of the $\pi_1(2015)$, observed by the E852 experiment \cite{E852:2004gpn,E852:2004rfa}, as the first radial excitation of the $\pi_1(1600)$.

\begin{table}[htbp]
\caption{The scale parameter $(\beta_\rho,~\beta_\lambda)$ of the SHO wave function for the low-lying light hybrid states (GeV). The first row is quantum numbers $(n_\lambda l_\lambda,~n_\rho l_\rho)$. The second and third rows are the concrete values of $\beta_\rho$ and $\beta_\lambda$ for the $q\bar{q}g$ and $s\bar{s}g$ states, respectively.} \label{table3}
\renewcommand\arraystretch{1.2}
\begin{tabular*}{86mm}{@{\extracolsep{\fill}}cccc}
\toprule[1pt]\toprule[1pt]
 $(00,~00)$       &  $(00,~01)$        &  $(01,~00)$          &   $(00,~10)$    \\
\toprule[1pt]
 $(0.242,~0.435)$ &  $(0.236,~0.415)$  &   $(0.227,~0.418)$   &    $(0.211,~0.378)$  \\
 $(0.268,~0.467)$ &  $(0.262,~0.446)$  &   $(0.251,~0.448)$   &    $(0.235,~0.407)$  \\
\bottomrule[1pt]\bottomrule[1pt]
\end{tabular*}
\end{table}

In addition to the average masses, we also determine the scale parameter $\beta$ of the SHO wave function by solving Eq. (\ref{eq12}). The specific values of $\beta_\lambda$ and $\beta_\rho$ are listed in Table \ref{table3}, which will be used to calculate the strong decays of $0^{+-}$ and $2^{+-}$ hybrids in Sec. \ref{sec3}.

\subsection{Further discussion of the masses of the hybrid states}\label{sec23}

In Sec. \ref{sec21}, we have introduced a transverse electric (TE) gluon to simulate the lowest excited gluonic field. To construct the TE gluon with quantum numbers $J_g^{P_gC_g} = 1^{+-}$, an intrinsic orbital angular momentum $L_g = 1$ is required \cite{Farina:2020slb}. The Clebsch-Gordan coefficient for the TE gluon is given by
\begin{equation}
\chi_{\mu,\nu}^{(-)} \equiv \langle s_g\nu,~L_g0|J_g\mu \rangle = \langle 1\nu,~10|1\mu \rangle = \frac{\lambda}{\sqrt{2}}\delta_{\mu\nu}. \label{eq16}
\end{equation}
In this framework, the orbital angular momentum $\textbf{\textit{L}}_H$ of a hybrid meson arises from the coupling of the excited gluonic field ($\textbf{\textit{J}}_g$) and the orbital angular momentum of the quark-antiquark pair ($\textbf{\textit{L}}_{q\bar{q}}$), i.e., $\textbf{\textit{L}}_H = \textbf{\textit{J}}_g + \textbf{\textit{L}}_{q\bar{q}}$. The total angular momentum $\textbf{\textit{J}}$ is then obtained by coupling $\textbf{\textit{L}}_H$ with the spin of the $q\bar{q}$ pair ($\textbf{\textit{S}}_{q\bar{q}}$). For clarity, the coupling order of the angular momenta is expressed as
\begin{equation}
\textbf{\textit{L}}_{q\bar{q}} \otimes \textbf{\textit{J}}_g \Rightarrow \textbf{\textit{L}}_H \otimes \textbf{\textit{S}}_{q\bar{q}} \Rightarrow \textbf{\textit{J}}. \label{eq17}
\end{equation}
The parity ($P$), charge conjugation ($C$), and $G$-parity of hybrid mesons containing a TE gluon are determined by
\begin{equation}
P = (-1)^{L_g + L_{q\bar{q}}}, \quad C = (-1)^{L_g + L_{q\bar{q}} + S_{q\bar{q}}}, \quad G = (-1)^{L_g + L_{q\bar{q}} + S_{q\bar{q}} + I}, \nonumber
\end{equation}
respectively. The allowed $I^GJ^{PC}$ quantum numbers and naming scheme for the low-lying light hybrid mesons are summarized in Table \ref{table4}.

\begin{table}[htbp]
\caption{The nomenclature and the allowed $I^GJ^{PC}$ quantum numbers for the ground and first orbitally excited hybrid mesons. States with the exotic $J^{PC}$ quantum
numbers are shown in bold.} \label{table4}
\renewcommand\arraystretch{1.2}
\begin{tabular*}{86mm}{@{\extracolsep{\fill}}cccccccc}
\toprule[1pt]\toprule[1pt]
 $l_{q\bar{q}}$ &    $j_g$ &    $L_H$    & $S_{q\bar{q}}$   & $I^G(J^{PC})$             &  meson     & $I^G(J^{PC})$             &  meson  \\
\toprule[1pt]
      0      &    1        &    1        &  0               & $1^+(1^{--})$             &  $\rho$    & $0^-(1^{--})$             &  $\omega,~\phi$ \\
      0      &    1        &    1        &  1               & $1^-(0^{-+})$             &  $\pi_0$   & $0^+(0^{-+})$             &  $\eta_0,~\eta^\prime_0$ \\
      0      &    1        &    1        &  1               & $\mathbf{1^-(1^{-+})}$    &  $\pi_1$   & $\mathbf{0^+(1^{-+})}$    &  $\eta_1,~\eta^\prime_1$ \\
      0      &    1        &    1        &  1               & $1^-(2^{-+})$             &  $\pi_2$   & $0^+(2^{-+})$             &  $\eta_2,~\eta^\prime_2$ \\
      1      &    1        &    0        &  0               & $1^-(0^{++})$             &  $a_0$     & $0^+(0^{++})$             &  $f_0,~f_0^\prime$   \\
      1      &    1        &    1        &  0               & $1^-(1^{++})$             &  $a_1$     & $0^+(1^{++})$             &  $f_1,~f_1^\prime$   \\
      1      &    1        &    2        &  0               & $1^-(2^{++})$             &  $a_2$     & $0^+(2^{++})$             &  $f_2,~f_2^\prime$   \\
      1      &    1        &    0        &  1               & $1^+(1^{+-})$             &  $b_{1h}$  & $0^-(1^{+-})$             &  $h_{1h},~h_{1h}^\prime$   \\
      1      &    1        &    1        &  1               & $\mathbf{1^+(0^{+-})}$    &  $b_0$     & $\mathbf{0^-(0^{+-})}$    &  $h_0,~h_0^\prime$   \\
      1      &    1        &    1        &  1               & $1^+(1^{+-})$             &  $b_{1l}$  & $0^-(1^{+-})$             &  $h_{1l},~h_{1l}^\prime$   \\
      1      &    1        &    1        &  1               & $\mathbf{1^+(2^{+-})}$    &  $b_{2l}$  & $\mathbf{0^-(2^{+-})}$    &  $h_{2l},~h_{2l}^\prime$   \\
      1      &    1        &    2        &  1               & $1^+(1^{+-})$             &  $b_{1h}$  & $0^-(1^{+-})$             &  $h_{1h},~h_{1h}^\prime$   \\
      1      &    1        &    2        &  1               & $\mathbf{1^+(2^{+-})}$    &  $b_{2h}$  & $\mathbf{0^-(2^{+-})}$    &  $h_{2h},~h_{2h}^\prime$   \\
      1      &    1        &    2        &  1               & $1^+(3^{+-})$             &  $b_3$     & $0^-(3^{+-})$             &  $h_3,~h_3^\prime$   \\

\bottomrule[1pt]\bottomrule[1pt]
\end{tabular*}
\end{table}

The low-lying light hybrid mesons with exotic quantum numbers have been denoted as bold in Table \ref{table4}. Among them, the $1^{-+}$ multiplet has been investigated in our previous work \cite{Chen:2023ukh}. In our scheme, the $0^{+-}$ and $2^{+-}$ states are the first orbitally excitations of $1^{-+}$ states, which is consistent with the results of lattice QCD. As argued in Refs. \cite{Dudek:2013yja,HadronSpectrum:2012gic}, the $0^{+-}$ and $2^{+-}$ hybrid states could be regarded as the systems which are composed of a $P$-wave colour-octet quark-antiquark pair coupled to a gluonic field with $J_g^{P_gC_g}=1^{+-}$.

When the SHO potential in the Eq. (\ref{eq4}) is replaced by the linear confinement potential and Coulomb potential, the spherical symmetry for the TE gluon\footnote{ In Sec. \ref{sec2}, the lowest excited gluonic field which is regarded as a TE gluon rotates around the center of $q\bar{q}$ pair (see Fig. \ref{hybrid} and Eq. (\ref{eq4})).} is broken while the cylindrical symmetry appears. Then the lowest excited gluonic field differentiates into two configurations, i.e., the $\Sigma_u^-$ and $\Pi_u$ gluonic fields. With the Born-Oppenheimer approximation, the results of lattice QCD indicates that the $\Sigma_u^-$ gluonic field may have higher excited energy \cite{Juge:2002br}. In Table \ref{table4}, the hybrid states with $L_H=0$ and $L_H=2$ may correspond to the $\Sigma_u^-$ field, while the states with $L_H=1$ correspond to the $\Pi_u$ field. To distinguish the states with the same $J$ but different $L_H$, we add a subscript ``$h$'' for the relative states with $L_H=0$ and $L_H=2$. Correspondingly, the states with $L_H=1$ are supplemented by a subscript ``$l$''.

At present, the mass pattern of light hybrid mesons is still unclear. What's more, this topic was rarely investigated by the phenomenological models in the literature. However, some valuable information could be obtained by the lattice QCD method \cite{Dudek:2013yja}, which are collected below.
\begin{enumerate}

\item The predicted masses of $\pi_1(1600)$ is about 1.2 times of the measured mass since the mass of light quark in the lattice QCD simulation is lager than physical value ($m_\pi=391$ MeV). With the measured mass of $\pi_1(1600)$ as benchmark, we may rescale the predicted masses of $0^{+-}$, $2^{+-}$, and another $1^{-+}$ states. We find that the average mass of one $0^{+-}$ and two $2^{+-}$ isovector states is about 2.10 GeV which is comparable to the our $(1P)_\rho$ mass in Table \ref{table2}.

\item The rescaled mass of an excited $1^{-+}$ hybrid is about 2.13 GeV, which is nearly equal to the average mass of $(2S)_\rho$ multiplet (see Table \ref{table2}). This result implies that the $\pi_1(2015)$ could be the first radially excited state of $\pi_1(1600)$ \cite{Meyer:2015eta}.

\item The mass of $b_{2h}$ is about 200 MeV above the average mass of $b_0$ and $b_{2l}$ state. Furthermore, the mass splitting of $b_0$ and $b_{2l}$ is about 100 MeV due to the probable spin-dependent interactions.

\item The light-strange mixing was found to be small for the isoscalar $0^{+-}$ and $2^{+-}$ hybrid mesons, which indicates the $h_0$, $h_{2l}$, and $h_{2h}$ are predominantly the $(u\bar{u}+d\bar{d})g/\sqrt{2}$ content while the $h_0^\prime$, $h_{2l}^\prime$, and $h_{2h}^\prime$ are predominantly the $s\bar{s}g$ content.
\end{enumerate}

Based on our result in Table \ref{table2} and  the conclusions above, we may further estimate the masses of $0^{+-}$ and $2^{+-}$ light hybrid mesons. The concrete results are listed in Table \ref{table5}, which will be taken to calculate the strong decays of $0^{+-}$ and
$2^{+-}$ hybrids in the next section.

\begin{table}[htbp]
\caption{The estimated masses of $0^{+-}$ and $2^{+-}$ light hybrid mesons (GeV).} \label{table5}
\renewcommand\arraystretch{1.2}
\begin{tabular*}{86mm}{@{\extracolsep{\fill}}cccccc}
\toprule[1pt]\toprule[1pt]
 $b_0/h_0$      &  $b_{2l}/h_{2l}$    &  $b_{2h}/h_{2h}$     &   $h_0^\prime$    &    $h_{2l}^\prime$  &  $h_{2h}^\prime$    \\
\toprule[0.8pt]
   1.94         &       2.04          &    2.18              &    2.10           &          2.20       &   2.39     \\
\bottomrule[1pt]\bottomrule[1pt]
\end{tabular*}
\end{table}

\section{The decay behaviors}\label{sec3}

The hybrid mass model developed in Sec. \ref{sec2} is consistent with the physical picture of the constituent gluon model \cite{Farina:2020slb}. This allows us to study both the mass spectrum and strong decays of hybrid mesons within a unified framework. Below, we give a brief review of the constituent gluon model for how to depict the decay process of a hybrid meson.

\begin{figure}[htbp]
\centering
\includegraphics[width=4.2cm,keepaspectratio]{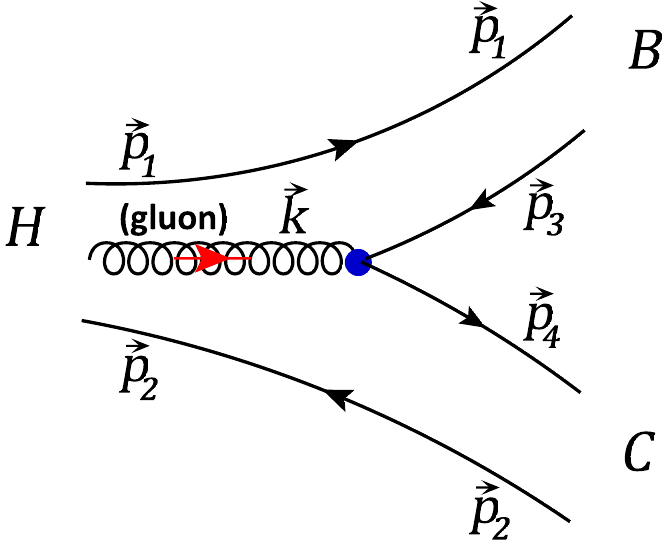}
\caption{Connected diagram for a hybrid state $H$ decaying into $B$ and $C$ mesons.}
\label{decay}
\end{figure}

For the decay process of a hybrid meson $H$ into mesons $B$ and $C$, the transition operator $\hat{\mathcal{T}}$ in the constituent gluon model can be expressed as
\begin{equation}
\begin{split}
\hat{\mathcal{T}} = &g_s \sum_{s,s^\prime,\lambda} \iiint \frac{\textup{d}^3\textbf{\textit{p}}_3\textup{d}^3\textbf{\textit{p}}_4\textup{d}^3\textbf{\textit{k}}}{\sqrt{2\omega_g}(2\pi)^6} \delta^3(\textbf{\textit{p}}_3 + \textbf{\textit{p}}_4 - \textbf{\textit{k}}) \\
& \times \frac{\lambda^{c_g}}{2} \phi_0^{(34)} \chi_s^\dagger \bm{\sigma} \tilde{\chi}_{s^\prime} \bm{\epsilon}(\hat{\textbf{k}},\nu) b_3^\dagger(\vec{p}_3) d_4^\dagger(\vec{p}_4) a_{\textbf{k}\nu}^{c_g}(\textbf{\textit{k}}),
\label{eq18}
\end{split}
\end{equation}
in the nonrelativistic limit. Here, $\lambda^{c_g}$ (${c_g} = 1, 2, \dots, 8$) are the Gell-Mann matrices, $\chi_{s^{(\prime)}}$ represent the quark spinors, and $\bm{\epsilon}^\mu(\hat{\textbf{k}},\nu)$ denotes the gluon polarization vector. Since the transition operator $\hat{\mathcal{T}}$ describes the dissociation of a constituent gluon into a quark-antiquark pair, the flavor wave function $\phi_0^{(34)}$ is taken as $(u\bar{u} + d\bar{d} + s\bar{s})/\sqrt{3}$. The parameter $\omega_g$, representing the effective mass of the constituent gluon \cite{Swanson:1997wy}, is set to 1.05 GeV in our calculations.

For calculating the helicity amplitude $\mathcal{M}^{j_H,j_B,j_C}(\textbf{\textit{p}})$ of a decay process $H \to B + C$ (see Fig. \ref{decay}), the spatial wave functions of hadrons could be described as the mock states \cite{Hayne:1981zy}. For a hybrid state $H$, the wave function is written as \cite{Chen:2023ukh}
\begin{equation}
\begin{split}
|H(&J_Hm_H[L_H(L_{q_1\bar{q}_2}J_g)S_{q_1\bar{q}_2}](\textbf{\textit{P}}_H)\rangle \equiv \\
& \omega_H \phi_H \Pi^{(H)}_{\textup{CG}} \int \textup{d}^3\textbf{\textit{p}}_1\textup{d}^3\textbf{\textit{p}}_2\textup{d}^3\textbf{\textit{k}} \delta^3(\textbf{\textit{p}}_1 + \textbf{\textit{p}}_2 + \textbf{\textit{k}} - \textbf{\textit{P}}_H) \\
& \times \Psi_{n_{q_1\bar{q}_2},n_g}^{L_{q\bar{q}}m_{q\bar{q}},L_gl_g}(\textbf{\textit{p}}_1,\textbf{\textit{p}}_2,\textbf{\textit{k}}) |q_1(\textbf{\textit{p}}_1)q_2(\textbf{\textit{p}}_2)g(\textbf{\textit{k}})\rangle. \label{eq19}
\end{split}
\end{equation}
The mock states for the final mesons $B$ and $C$ could be found in our previous work \cite{Chen:2016iyi}. In Eq. (\ref{eq19}), $\omega_H$ and $\phi_H$ denote the color and flavor wave functions of the initial hybrid state, respectively. As a color singlet, the color wave function of the hybrid state is $\omega_H = |[(q_1\bar{q}_2)_8 \otimes g_8]_1\rangle$. The flavor wave function of the hybrid state is identical to that of a meson state, since the gluon is flavorless. The term $\Pi^{(H)}_{\textup{CG}}$ in Eq.~(\ref{eq19}) involves the product of Clebsch-Gordan coefficients, representing the coupling of quark-gluon spins and angular momenta to the total spin $J_H$ of the hybrid state. Specifically, it is given by
\begin{equation}
\begin{split}
\Pi^{(H)}_{\textup{CG}} = & \sqrt{\frac{2J_g + 1}{4\pi}} \mathcal{D}^{J_g\ast}_{m_g,\mu}(\hat{\textbf{k}}) \chi^{(-)}_{\mu,\nu} \langle l_{q_1\bar{q}_2}m_l, J_gm_g | L_Hm_L \rangle \\
& \times \langle \frac{1}{2}m_1, \frac{1}{2}m_2 | S_Hs_H \rangle \langle S_Hs_H, L_Hm_L | J_Hm_H \rangle. \label{eq20}
\end{split}
\end{equation}
Here, $\chi^{(-)}_{\mu,\nu}$ is defined in Eq. (\ref{eq16}). The rotation matrix $\mathcal{D}^{J_g\ast}_{m_g,\mu}$ in Eq.~(\ref{eq20}) transforms the gluon's angular momentum projection into the basis of hybrid system. For simplification, we use the relation \cite{Farina:2020slb}
\begin{equation}
\chi_s^\dagger \bm{\sigma} \tilde{\chi}_{s^\prime} \bm{\epsilon}(\hat{\textbf{k}},\nu) = -2^{|s + s^\prime|/2} \mathcal{D}^1_{s + s^\prime,\nu}(\hat{\textbf{k}}), \label{eq21}
\end{equation}
in practical calculations. Combining the two "$\mathcal{D}$" functions in Eqs. (\ref{eq20}) and (\ref{eq21}), we obtain
\begin{equation}
\begin{split}
\sqrt{\frac{2J_g + 1}{4\pi}} & \mathcal{D}^{J_g\ast}_{m_g,\mu}(\hat{\textbf{k}}) \left( -2^{|s + s^\prime|/2} \mathcal{D}^1_{s + s^\prime,\nu}(\hat{\textbf{k}}) \right) = (-1)^{m_g + 1} \\
& \times 2^{(|s + s^\prime| - 1)/2} \langle J_g- m_g, 1s + s^\prime |L_gl_g \rangle \mathcal{Y}^\ast_{L_gl_g}(\hat{\textbf{k}}), \label{eq22}
\end{split}
\end{equation}
for the decay process of a hybrid state with a TE gluon. For the $1^{+-}$ gluon, $J_g=L_g=1$ and $l_g=s+s^\prime-m_g$ are implied. As shown in Eq. (\ref{eq5}), the spatial wave function of a hybrid state can be approximated as a simple product of $\rho$- and $\lambda$-modes. In momentum space, the spatial wave functions of the $0^{+-}$ and $2^{+-}$ states are denoted as
\begin{equation}
\Psi^{1m_{q\bar{q}},1l_g}_{0,0}(\textbf{\textit{p}}_\rho,\textbf{\textit{p}}_\lambda) = \frac{8/3}{\pi^{1/2}\beta_\rho^{5/2}\beta_\lambda^{5/2}} \mathcal{Y}_{1m_{q\bar{q}}}(\textbf{\textit{p}}_\rho) \mathcal{Y}_{1l_g}(\textbf{\textit{p}}_\lambda) \textup{e}^{-\frac{p^2_\rho}{2\beta^2_\rho} - \frac{p^2_\lambda}{2\beta^2_\lambda}}. \label{eq23}
\end{equation}
Using Jacobi coordinates, $\textbf{\textit{p}}_\rho$ and $\textbf{\textit{p}}_\lambda$ can be simplified as
\begin{equation}
\textbf{\textit{p}}_\rho = \frac{\textbf{\textit{q}}}{2} - \textbf{\textit{p}}; \quad \textbf{\textit{p}}_\lambda = \textbf{\textit{k}}, \label{eq24}
\end{equation}
in the rest frame of the light $q\bar{q}g$ hybrid system. Here, $\textbf{\textit{p}}$ is the three-momentum of the daughter meson $B$ in the center-of-mass frame of $H$ (see Fig. \ref{decay}), while $\textbf{\textit{q}}$ and $\textbf{\textit{k}}$ are the integration variables. The partial wave amplitude is given by

\begin{widetext}
\begin{equation}
\begin{aligned}\label{eq25}
\mathcal{M}^{H \to BC}_{LS}(p) = g_s & \frac{\sqrt{2L + 1}}{2J_H + 1} \sum_{\text{$j_B$,$j_C$}} \langle L0Jj_H | J_Hj_A \rangle \langle J_Bj_B, J_Cj_C | Jj_H \rangle \langle l_{q_1\bar{q}_2}m_l, J_gm_g | L_Hm_L \rangle \langle \frac{1}{2}m_1, \frac{1}{2}m_2 | S_Hs_H \rangle \langle S_Hs_H, L_Hm_L | J_Hm_H \rangle \\
& \times \langle \frac{1}{2}m_1, \frac{1}{2}s^\prime | S_Bs_B \rangle \langle S_Bm_B, L_Bm_{L_B} | J_Bm_{JB} \rangle \langle \frac{1}{2}m_2, \frac{1}{2}s | S_Cm_C \rangle \langle S_Cm_C, L_Cm_{L_C} | J_Cm_{J_C} \rangle 2^{|s + s^\prime|/2} (-1)^{s + s^\prime} \chi^{(-)}_{\mu,\nu} \\
& \times \langle 1 - m_g, 1s + s^\prime | 1l_g \rangle \xi_c \xi_f \iint \textup{d}^3\textbf{\textit{q}}\textup{d}^3\textbf{\textit{k}} ~ \psi^{L_{q\bar{q}}m_{q\bar{q}}}_{n_{q\bar{q}}}\left(\frac{\textbf{\textit{q}}}{2} - \textbf{\textit{p}}\right) \psi^{L_gl_g}_{n_g}\left(\textbf{\textit{k}}\right) \psi^{\ast}_B\left(\frac{\textbf{\textit{k}} + \textbf{\textit{q}}}{2} - \varepsilon\textbf{\textit{p}}\right) \psi^{\ast}_C\left(\frac{\textbf{\textit{k}} - \textbf{\textit{q}}}{2} + \varepsilon\textbf{\textit{p}}\right).
\end{aligned}
\end{equation}
\end{widetext}

The color coefficient $\xi_c$ is determined by the matrix element $\xi_c = \langle (q_1\bar{q}_3)_1 (\bar{q}_2q_4)_1 | [(q_1\bar{q}_2)_8 \otimes g_8]_1 \rangle = \sqrt{8/9}$. The flavor coefficient $\xi_f$ is calculated with the same method as for the ordinary $q\bar{q}$ mesons, since the gluon is flavorless. The parameter $\varepsilon$ is defined as $\varepsilon=m^\prime/(m+m^\prime)$, where the $m$ and $m^\prime$ denote the masses of 1 and 3 quarks in Fig. \ref{decay}, respectively. The partial width of the decay process $H \to BC$ in the rest frame of $H$ is given by
\begin{equation}
\begin{aligned}\label{eq26}
\Gamma(H \rightarrow BC) = 2\pi \frac{E_B E_C}{M_H} p \sum_{L,S} |\mathcal{M}^{H \to BC}_{LS}(p)|^2.
\end{aligned}
\end{equation}

In our calculations, the scale parameters $\beta$ of SHO wave function for the final mesons are also obtained by the potential model (see Eq. (\ref{eq15})). The masses of the final mesons are taken from the Particle Data Group (PDG) \cite{ParticleDataGroup:2024cfk}. For the undiscovered mesons, we fix their masses using the potential model. The strong coupling constant $g_s$ is determined by the decay width of $\pi_1(1600)$ under the $1^{-+}$ hybrid assignment.


As discussed in Sec. \ref{sec2}, lattice QCD results indicate that one isoscalar hybrid in a light-flavor nonet is predominantly a $(u\bar{u} + d\bar{d})g/\sqrt{2}$ state, while the other is predominantly an $s\bar{s}g$ state. For simplicity, we initially ignore flavor mixing between the two isoscalar hybrids. Mixing may also occur between two $2^{+-}$ hybrid mesons with different $L_H$ (see Table \ref{table4}), such as the $b_{2l}$ and $b_{2h}$ states. However, the mixing angles for these states cannot be quantitatively analyzed in our current framework. Nevertheless, we will discuss the potential effects of mixing on their decay behaviors. To this end, the nine hybrid states in Table \ref{table5} are divided into four groups: $(b_0, h_0, h^\prime_0)$, $(b_{2l}, b_{2h})$, $(h_{2l}, h_{2h})$, and $(h^\prime_{2l}, h^\prime_{2h})$. We now discuss their decay properties in detail.

\subsection{The $b_0$,~$h_0$,~and $h^\prime_0$ states}\label{sec31}

The allowed decay channels of the $b_0$, $h_0$, and $h^\prime_0$ states are listed in Table \ref{table6}, along with their predicted decay widths. The $b_0(1940)$ and $h^\prime_0(2100)$ are expected to be broad, primarily due to the $s$-wave decay channel ``$0^{+-} \to 2^1S_0 + 1^1S_0$''. The broad nature of the $b_0(1940)$ and $h^\prime_0(2100)$ states poses a significant challenge for the experimental searches in the future.
The $h_1(1170)\pi$ channel also contributes significantly for the total decay width of $b_0(1940)$ state. Additionally, the $b_1(1235)\eta$ and $K_1(1270)K$ channels are likely to be observable.

\begin{table}[htbp]
\caption{The partial and total widths of the  $b_0$, $h_0$, and $h^\prime_0$ states with $J^{PC}=0^{+-}$ (MeV). When the partial width of a decay channel is predicted to be smaller than 0.1 MeV, the result is denoted as ``$0.0$''. The uncertainty which exists in our results is about 10\%$\sim$30\%.} \label{table6}
\renewcommand\arraystretch{1.2}
\begin{tabular*}{86mm}{@{\extracolsep{\fill}}lclclc}
\toprule[1pt]\toprule[1pt]
 $b_0(1940)$      & $\Gamma_i$       & $h_0(1940)$     & $\Gamma_i$      & $h^\prime_0(2100)$       & $\Gamma_i$      \\
\cline{1-2}\cline{3-4}\cline{5-6}
 $h_1(1170)\pi$   & 26.3             & $b_1(1235)\pi$  & 66.8            & $h^\prime_1(1415)\eta$   & 2.6   \\
 $b_1(1235)\eta$  & 5.0              & $h_1(1170)\eta$ & 7.5             & $K_1(1270)K$             & 33.7  \\
 $K_1(1270)K$     & 4.2              & $K_1(1270)K$    & 1.6             & $K_1(1400)K$             & 53.3   \\
 $K_1(1400)K$     & 1.0              & $K_1(1400)K$    & 1.1             & $K(1460)K$               & 394.4    \\
 $\omega_2\pi$    & 0.0              & $\rho_2\pi$     & 0.0             &                          &         \\
 $\pi(2S)\pi$     & 267.1            &                 &                 &                          &        \\
\cline{1-2}\cline{3-4}\cline{5-6}
 Total            & 303.6            & Total           &  77.0           &  Total                   & 484.0   \\
\bottomrule[1pt]\bottomrule[1pt]
\end{tabular*}
\end{table}

The $h_0(1940)$ appears to be relatively narrow if its mass lies below the threshold of the $K^\ast(1460)K$ channel. Its dominant decay mode is $b_1(1235)\pi$, but it could also be observed in the $h_1(1170)\eta$, $K_1(1270)K$, and $K_1(1400)K$ channels. However, if the $h_0$ state is above the $K^\ast(1460)K$ threshold, it could become significantly broad. For instance, if the mass of $h_0$ exceeds the $K^\ast(1460)K$ threshold by about 40 MeV, the partial width for $h_0 \to K^\ast(1460)K$ could exceed 100 MeV.

For the $h^\prime_0(2100)$, the $K_1(1270)K$, $K_1(1400)K$, and $K(1460)K$ channels are the main decay modes, while the partial width for $h^\prime_1(1415)\eta$ is relatively small. These findings for the $b_0$, $h_0$, and $h^\prime_0$ states are consistent with the results from Ref. \cite{Page:1998gz}.

We should give some comments for the uncertainties for the predicted partial and total widths which are listed in Table \ref{table6} and the following tables. We check our results by changing the parameter values (such as the $m_q$ and $m_g$) within a reasonable range. The predicted widths are found to fluctuate around 10\%. However, the uncertainties may also arise from the strong decay constant $g_s$ due to the inaccurately measured decay width of $\pi_1(1600)$. Based on the considerations above, we conclude that the uncertainty of our results in Table \ref{table6}-\ref{table9} is about 10\%$\sim$30\%. The same comment has been given in our previous work \cite{Chen:2023ukh}.

\subsection{The $b_{2l}$ and $b_{2h}$ states}\label{sec32}

\begin{table}[htbp]
\caption{The partial and total widths of the $b_{2l}$ and $b_{2h}$ states with $J^{PC}=2^{+-}$ (MeV). If the hybrid state is below the threshold of a decay channel, the result is denoted as ``$-$''.} \label{table7}
\renewcommand\arraystretch{1.2}
\begin{tabular*}{86mm}{@{\extracolsep{\fill}}cccccc}
\toprule[1pt]\toprule[1pt]
 channel          & $b_{2l}(2040)$   & $b_{2h}(2190)$  &                   &                &       \\
\cline{1-3}
 $\omega\pi$      & 0.1              & 0.1             & $\pi(2S)\pi$      & 0.4            &  3.0     \\
 $\rho\eta$       & 0.0              & 0.0             & $K(1460)K$        & 0.0            &  0.1     \\
$\rho\eta^\prime$ & 0.0              & 0.0             & $K^\ast(1410)K$   & 0.0            &  0.1    \\
 $K^\ast K$       & 0.0              & 0.1             & $\omega(2S)\pi$   & 0.7            &  0.6       \\
 $h_1(1170)\pi$   & 4.3              & 12.9            & $\rho(2S)\eta$    & 0.0            &  0.1       \\
 $b_1(1235)\eta$  & 1.4              & 5.9             & $\pi(3S)\pi$      & 0.4            &  2.6        \\
 $a_1(1260)\pi$   & 9.8              & 12.2            & $f_1(1280)\rho$   & $-$            &  12.8       \\
 $a_2(1320)\pi$   & 7.4              & 0.2             & $f_2(1270)\rho$   & $-$            &  9.9        \\
 $K_1(1270)K$     & 0.8              & 3.4             & $\omega_1(1D)\pi$ & 0.0            &  0.1        \\
 $K_1(1400)K$     & 2.5              & 16.6            & $\omega_2(1D)\pi$ & 17.2           &  7.8        \\
$K_2^\ast(1430)K$ & 1.6              & 1.2             & $\omega_3(1D)\pi$ & 0.1            &  0.0        \\
\cline{4-6}
                  &                  &                 &  Total            & 46.7           &  89.7          \\
\bottomrule[1pt]\bottomrule[1pt]
\end{tabular*}
\end{table}

The $b_{2l}(2040)$ and $b_{2h}(2190)$ states are predicted to be significantly narrower than the $b_0(1940)$. In particular, the total width of the $b_{2l}(2040)$ is estimated to be 46.7 MeV. Promising decay channels for the $b_{2l}(2040)$ include $h_1(1170)\pi$, $a_1(1260)\pi$, $a_2(1320)\pi$, $K_1(1400)K$, and $\omega_2(1D)\pi$, making them ideal for the experimental searches. Ref. \cite{Page:1998gz} also suggested the $b_{2l}$ to be a narrow state, with a predicted total width of only 11 MeV when its mass was taken as 2.0 GeV. The $h_1(1170)\pi$ and $a_1(1260)\pi$ channels were identified as the dominant decay modes for the $b_2$ hybrid state, which aligns with our findings in Table \ref{table7}. However, the flux-tube model \cite{Close:1994hc} predicts a much broader total width of 200--400 MeV for a $b_2$ state. Despite this discrepancy, the flux-tube model also supports the $h_1(1170)\pi$, $a_1(1260)\pi$, and $a_2(1320)\pi$ channels as the primary decay modes for the $b_2$ state.

The total decay width of $b_{2h}(2190)$ state, with $L_H = 2$, is predicted to be 89.7 MeV. This state could be observed in the $h_1(1170)\pi$, $a_1(1260)\pi$, $K_1(1400)K$, $f_1(1280)\rho$, $f_2(1270)\rho$, and $\omega_2(1D)\pi$ channels. The decay patterns of the $b_{2h}(2190)$ are similar to those of the $b_{2l}(2040)$, making it challenging to distinguish their mixing effects. However, the $a_2(1320)\pi$ channel is particularly noteworthy: it is a dominant decay mode for the $b_{2l}(2040)$, but contributes minimally to the $b_{2h}(2190)$. If future experiments discover two $2^{+-}$ isovector states in the $a_2(1320)\pi$ channel, which may indicate an evidence of mixing between the $b_{2l}$ and $b_{2h}$ states.

\subsection{The $h_{2l}$ and $h_{2h}$ states}\label{sec33}

\begin{table}[htbp]
\caption{The partial and total widths of the $h_{2l}$ and $h_{2h}$ states with $J^{PC}=2^{+-}$ (MeV).} \label{table8}
\renewcommand\arraystretch{1.2}
\begin{tabular*}{86mm}{@{\extracolsep{\fill}}cccccc}
\toprule[1pt]\toprule[1pt]
 channel            & $h_{2l}(2040)$   & $h_{2h}(2190)$  &                    &                &       \\
\cline{1-3}
 $\rho\pi$          & 0.2              & 0.0             & $b_1(1235)\pi$     & 11.4           &  35.8    \\
 $\omega\eta$       & 0.0              & 0.1             & $a_1(1260)\rho$    & 14.1           &  59.6    \\
$\omega\eta^\prime$ & 0.0              & 0.0             & $a_2(1320)\rho$    & $-$            &  17.1    \\
 $K^\ast K$         & 0.0              & 0.1             & $K_1(1270)K$       & 0.3            &  2.9       \\
 $\rho(1450)\pi$    & 1.3              & 1.4             & $K_1(1400)K$       & 1.7            &  7.7     \\
 $\omega(1420)\eta$ & 0.0              & 0.1             & $K_2^\ast(1430)K$  & 0.7            &  0.1     \\
  $K(1460)K$        & 0.0              & 0.2             & $\rho_1(1D)\pi$    & 0.0            &  0.2       \\
 $K^\ast(1410)K$    & 0.0              & 0.1             & $\rho_2(1D)\pi$    & 51.9           &  23.3       \\
 $h_1(1170)\eta$    & 1.8              & 6.6             & $\rho_3(1D)\pi$    & 0.1            &  0.0       \\
\cline{4-6}
                    &                  &                 &  Total             & 83.5           &  155.3     \\
\bottomrule[1pt]\bottomrule[1pt]
\end{tabular*}
\end{table}

As the light isoscalar partners of the $b_{2l}$, the $h_{2l}$ is expected to have nearly equal masses, as it is a predominant $(u\bar{u} + d\bar{d})g/\sqrt{2}$ state. The case of  $h_{2h}$ and $b_{2h}$ states is alike. The decay properties of $h_{2l}$ and $h_{2h}$ have been investigated, and the results are summarized in Table \ref{table8}.

The $b_1(1235)\pi$, $a_1(1260)\rho$, and $\rho_2(1D)\pi$ channels are predicted to be the dominant decay modes for the $h_{2l}$ state, with a total decay width of 83.5 MeV. Ref. \cite{Page:1998gz} also supports the $b_1(1235)\pi$ as a primary decay channel for the $h_{2l}$ state. Interestingly, the E852 experiment at Brookhaven National Laboratory (BNL) has reported a probable $h_{2l}$ state around 2.0 GeV in 2005 \cite{Adams:2005tx}. Specifically, a strong signal of a resonance structure with exotic quantum numbers $I^GJ^{PC} = 0^-2^{+-}$ was observed in the process $\pi^- + p \to h_2(0^-2^{+-}) + n \to b_1\pi n \to \omega\pi\pi n \to 5\pi n$. The properties of this $h_2$ state align well with the $h_{2l}$ candidate in our framework. However, further experimental evidence is needed to confirm the existence of this state.

The total decay width of $h_{2h}(2190)$ state, with $L_H = 2$, is predicted to be 155.3 MeV. This state could be observed in the decay channels such as the $h_1(1170)\eta$, $b_1(1235)\pi$, $a_1(1260)\rho$, $a_2(1320)\rho$, $K_1(1400)K$, and $\rho_2(1D)\pi$. The decay patterns of the $b_{2h}(2190)$ are similar to those of the $b_{2l}(2040)$, making it challenging to distinguish their mixing effects based solely on the decay behaviors.

\subsection{The $h^\prime_{2l}$ and $h^\prime_{2h}$ states}\label{sec34}

\begin{table}[htbp]
\caption{The partial and total widths of the $h^\prime_{2l}$ and $h^\prime_{2h}$ states with $J^{PC}=2^{+-}$ (MeV).} \label{table9}
\renewcommand\arraystretch{1.2}
\begin{tabular*}{86mm}{@{\extracolsep{\fill}}cccccc}
\toprule[1pt]\toprule[1pt]
 channel                & $h^\prime_{2l}(2200)$   & $h^\prime_{2h}(2390)$  &      &                &       \\
\cline{1-3}
 $K^\ast K$             & 0.1              & 0.0             & $K(1460)K$         & 0.3            &  6.2   \\
 $\phi\eta$             & 0.0              & 0.0             & $K^\ast(1410)K$    & 2.8            &  3.4    \\
 $\phi\eta^\prime$      & 0.0              & 0.0             & $\phi(1680)\eta$   & $-$            &  0.2    \\
 $h_1^\prime(1415)\eta$ & 0.7              & 3.6             & $K^\ast(1680)K$    & $-$            &  0.3       \\
 $K_1(1270)K$           & 2.3              & 14.5            & $K_2(1770)K$       & $-$            &  19.1     \\
 $K_1(1400)K$           & 18.1             & 47.6            & $K_2^\ast(1780)K$  & $-$            &  0.0     \\
 $K_2^\ast(1430)K$      & 9.4              & 1.1             &                    &                &          \\
\cline{4-6}
                        &                  &                 &  Total             & 33.7           &  96.0    \\
\bottomrule[1pt]\bottomrule[1pt]
\end{tabular*}
\end{table}

The strong decays of the $h^\prime_{2l}$ and $h^\prime_{2h}$ states are predicted in Table \ref{table9}. According to our results, the $h^\prime_{2l}$ is likely to be a narrow hybrid state, predominantly decaying into the $K_1(1270)K$ and $K_2^\ast(1430)K$ channels. These findings are consistent with the results of Ref. \cite{Page:1998gz}.

In contrast, the $h^\prime_{2h}$ state is predicted to be broader. Its main decay channels include the $K_1(1270)K$, $K_1(1400)K$, and $K_2(1770)K$. Investigating the $h^\prime_{2l}$ and $h^\prime_{2h}$ states in the $K_2^\ast(1430)K$ channel could provide insights into their mixing effects, which is similar to the case of the $b_{2l}$ and $b_{2h}$ states. We particularly emphasize the $K_2(1770)K$ channel for the $h^\prime_{2h}$ state. As an $s$-wave decay channel, its partial width is predicted to be 19.1 MeV. Similar results are also observed for the $b_{2l}$, $b_{2h}$, $h_{2l}$, and $h_{2h}$ states (see Tables \ref{table7} and \ref{table8}).

\section{Conclusion and outlook}\label{sec4}

Investigating hybrid mesons with the exotic $J^{PC}$ quantum numbers is a crucial step toward establishing the complete spectrum of hybrid states. Since these hybrid mesons cannot mix with conventional $q\bar{q}$ mesons, their properties are easier to analyze. Currently, only the $\pi_1(1600)$ and $\eta_1(1855)$ with $J^{PC} = 1^{-+}$ have been definitively observed. However, the probable evidences for states with $J^{PC} = 2^{+-}$ and $J^{PC} = 3^{-+}$ have also been detected in previous experiments \cite{Adams:2005tx,COMPASS:2014vkj}, indicating the potential for discovering more hybrid mesons with exotic quantum numbers in the future.

In this work, we focus on the mass spectrum and strong decays of hybrid mesons with $J^{PC} = 0^{+-}$ and $J^{PC} = 2^{+-}$. We developed a constituent gluon model in which the lowest excited gluonic field is approximated as a transverse electric (TE) gluon. In this scheme, the $0^{+-}$ and $2^{+-}$ hybrids can be interpreted as orbital excitations of the $1^{-+}$ hybrids. To describe the masses of hybrid mesons, we assume that the long-range confinement potential exists only between the quark and the TE gluon. Initially, we used a simple harmonic oscillator (SHO) potential to test the model. By solving the three-body Schr\"{o}dinger equation, we found that the predicted excitation energies of the $\lambda$-mode are consistent with the results from other methods when interpreted as gluelump states. We then replaced the SHO potential with a linear confinement potential and included a Coulomb potential between the quark-antiquark pair. The resulting mass predictions for the $0^{+-}$ and $2^{+-}$ light hybrid states agree with lattice QCD results \cite{Dudek:2013yja}. The hybrid potential model developed in Sec. \ref{sec2} is fully compatible with the constituent gluon model \cite{Farina:2020slb}.

In Sec. \ref{sec3}, we systematically investigated the strong decays of $0^{+-}$ and $2^{+-}$ light hybrid mesons using the constituent gluon model. The main decay properties of these hybrids are summarized as follows:

\begin{enumerate}
    \item The $b_0$ and $h_0^\prime$ states are likely to be broad, with predicted total decay widths exceeding 300 MeV (see Table \ref{table6}). If the $h_0$ lies below the threshold of the $K(1460)K$ channel (an $s$-wave decay channel), its total width is predicted to be around 80 MeV. However, if the $K(1460)K$ channel is accessible, the $h_0$ could also become a broad state. This makes the experimental search for $0^{+-}$ light hybrids particularly challenging.

    \item The $b_{2l}$ and $h_{2l}$ states are expected to be relatively narrow, with predicted widths of less than 100 MeV. The dominant decay modes for the $b_{2l}$ state are $a_1(1260)\pi$, $a_2(1320)\pi$, and $\omega_2(1D)\pi$. As a $s$-wave decay, the $\rho_2(1D)\pi$ channel has the largest partial width for the $h_{2l}$ state, while the $b_1(1235)\pi$ and $a_1(1260)\rho$ channels also play significant roles.

    \item The $b_{2h}$ and $h_{2h}$ states are predicted to have total decay widths of approximately 90$-$150 MeV. The $b_{2h}$ state primarily decays into $h_1(1170)\pi$, $a_1(1260)\pi$, $K_1(1400)K$, $f_1(1280)\rho$, and $f_2(1270)\rho$ channels. As the isoscalar partner of $b_{2h}$, the $h_{2h}$ mainly decays into $b_1(1235)\pi$, $a_1(1260)\rho$, $a_2(1320)\rho$, and $\rho_2(1D)\pi$ channels.

    \item The $h^\prime_{2l}$ state is expected to be a narrow resonance, with a predicted decay width of only 33.7 MeV. Experimental searches for this state could focus on the $K_1(1400)K$ and $K^\ast_2(1430)K$ channels. The $h^\prime_{2h}$ state, with a predicted width of 96.0 MeV, primarily decays into $K_1(1270)K$, $K_1(1400)K$, and $K_2(1770)K$ channels.
\end{enumerate}

In a word, we have studied the masses and strong decays of $0^{+-}$ and $2^{+-}$ hybrid mesons using an unified set of model parameters. We suggest potential experimental processes for discovering these states. Since signals of $2^{+-}$ and $3^{-+}$ hybrids have been observed in $\pi^-p$ inelastic scattering, the COMPASS experiment is well-positioned to discover $0^{+-}$ and $2^{+-}$ hybrid states in the near future. Although $J/\psi$ radiative decays are forbidden for $0^{+-}$ and $2^{+-}$ states due to $C$-parity conservation, the BESIII collaboration could still search for these states. For example, the $\pi_1(1600)$ was observed in the cascade process $\chi_{c1} \to \pi_1(1600)\pi \to \eta^\prime\pi\pi$ \cite{Liu}. Similarly, the $b_{2l}$ and $b_{2h}$ states could be discovered in the processes $\chi_{c1} \to \rho + b_{2}$ ($p$-wave) and $J/\psi \to \pi + b_{2}$ ($d$-wave), while the $h_{2l}$ and $h_{2h}$ states could be found in $\chi_{c1} \to \omega + h_{2}$ ($p$-wave).

\section*{Acknowledgement}

B. C. thanks Professor F. Sch\"{o}ber for providing the details of their previous work \cite{Schoberl:1984ha}. This project is supported by the National Natural Science Foundation of China under Grant No. 12335001 and No. 12247101, the `111 Center' under Grant No. B20063, the Natural Science Foundation of Gansu Province (No. 22JR5RA389), the fundamental Research Funds for the Central Universities, and the project for top-notch innovative talents of Gansu province. B. C. is also supported by the key Discipline Construction Project of Anhui Science and Technology University (No. XK-XJGY002).

\appendix
\section{Solving Eq. (\ref{eq12}) by variational method}\label{A}

Here, we provide details on solving the three-body Schr\"{o}dinger equation, i.e., Eq. (\ref{eq12}), using the variational method. We focus on the ground state to illustrate how the expectation value of the Hamiltonian in Eq. (\ref{eq12}) is obtained. The final result is given by Eq. (\ref{eq14}).

First, we express the wave function of a ground hybrid state in momentum space as
\begin{equation}
\Psi_{00,00}(\textbf{\textit{p}}_\rho,\textbf{\textit{p}}_\lambda) = \frac{1}{\pi^{3/2}\beta_\rho^{3/2}\beta_\lambda^{3/2}} \textup{e}^{ -\frac{p^2_\rho}{2\beta^2_\rho} - \frac{p^2_\lambda}{2\beta^2_\lambda} }, \label{A1}
\end{equation}
which corresponds to Eq. (\ref{eq13}) in configuration space. We define the following Jacobi coordinates (see Fig. \ref{hybrid}):
\begin{equation}
\bm{\rho} = \bm{r}_q - \bm{r}_{\bar{q}}, \quad \bm{\lambda} = \bm{r}_g - \frac{\bm{r}_q + \bm{r}_{\bar{q}}}{2}. \label{A2}
\end{equation}
The expectation values of the following terms in the Hamiltonian of Eq. (\ref{eq12}) can be calculated directly:
\begin{equation}
\langle\Psi_{00,00}|\frac{\textbf{p}_\rho^2}{2m_\rho} + \frac{\textbf{p}_\lambda^2}{2m_\lambda} + \frac{1}{6}\frac{\alpha_s}{\rho}|\Psi_{00,00}\rangle = \frac{3}{4} \left( \frac{\beta_\rho^2}{m_\rho} + \frac{\beta_\lambda^2}{m_\lambda} \right) + \frac{\alpha_s \beta_\rho}{3\sqrt{\pi}}. \nonumber
\end{equation}
Calculating the expectation value of the term ``$9b(r_1 + r_2)/8$'' in Eq. (\ref{eq12}) requires additional techniques. To this end, we define the following Jacobi coordinates:
\begin{equation}
\bm{\rho}^\prime = \bm{r}_1 = \bm{r}_g - \bm{r}_q, \quad \bm{\lambda}^\prime = \bm{r}_{\bar{q}} - \frac{m_q\bm{r}_q + m_g\bm{r}_g}{m_q + m_g}. \label{A3}
\end{equation}
The transformation between $(\bm{\rho}, \bm{\lambda})$ and $(\bm{\rho}^\prime, \bm{\lambda}^\prime)$ is given by
\begin{eqnarray}\label{A4}
\begin{aligned}
 \left(
           \begin{array}{c}
                     \bm{\rho}   \\
                     \bm{\lambda}\\
                    \end{array}
     \right) &= \left(
           \begin{array}{cc}
                    -\epsilon              & -1           \\
                    1 - \frac{\epsilon}{2} & -\frac{1}{2} \\
                    \end{array}
     \right)  \left(
           \begin{array}{c}
                     \bm{\rho}^\prime \\
                     \bm{\lambda}^\prime \\
                    \end{array}
     \right),
\end{aligned}
\end{eqnarray}
where $\epsilon = \frac{m_g}{m_q + m_g}$. The determinant of the matrix in Eq. (\ref{A4}) is equal to 1, ensuring that ``$\iint\textup{d}^3\bm{\rho}\textup{d}^3\bm{\lambda} = \iint\textup{d}^3\bm{\rho}^\prime\textup{d}^3\bm{\lambda}^\prime$''. Using Eq. (\ref{A4}), the wave function of the ground hybrid state, i.e., Eq. (\ref{eq13}), can be rewritten as
\begin{equation}
\Psi_{00,00}(\bm{\rho}^\prime,\bm{\lambda}^\prime) = \frac{\beta_\rho^{3/2}\beta_\lambda^{3/2}}{\pi^{3/2}} \textup{e}^{ - A\rho^{\prime2} + B\bm{\rho}^\prime \cdot \bm{\lambda}^\prime - C \lambda^{\prime2} }.  \label{A5}
\end{equation}
Here, we define
\begin{equation}
A = \frac{1}{2}\left( \epsilon^2\beta^2_\rho + \xi^2\beta_\lambda^2 \right), \quad B = -\epsilon\beta^2_\rho + \frac{\xi}{2}\beta_\lambda^2, \quad C = \frac{\beta^2_\rho}{2} + \frac{\beta_\lambda^2}{8}, \nonumber
\end{equation}
where $\xi = 1 - \frac{\epsilon}{2}$. The expectation value
\begin{equation}
\langle\Psi_{00,00}(\bm{\rho}^\prime,\bm{\lambda}^\prime)| \frac{9b}{8}r_1 |\Psi_{00,00}(\bm{\rho}^\prime,\bm{\lambda}^\prime)\rangle, \nonumber
\end{equation}
can now be calculated directly. The expectation value of ``$9br_2/8$'' is obtained similarly. Combining these results, the energy of the ground state (Eq. (\ref{eq14})) is derived. One can also calculate the energy of the $1\rho$ excited states, yielding
\begin{equation}
E_{01,00} = \frac{5\beta_\rho^2}{4m_\rho} + \frac{3\beta_\lambda^2}{4m_\lambda} + \frac{3b\left(\beta_\lambda^2+3\beta_\rho^2\right)}{2\sqrt{\pi}\beta_\rho\beta_\lambda\sqrt{\beta_\lambda^2+4\beta_\rho^2}} + \frac{2\alpha_s \beta_\rho}{9\sqrt{\pi}} + c^\prime_H.   \nonumber
\end{equation}

%



\begin{thebibliography}{99}

\bibitem{Gross:2022hyw}
  F.~Gross, E.~Klempt, S.~J.~Brodsky, \textit{et al.},
  50 Years of Quantum Chromodynamics,
  Eur. Phys. J. C \textbf{83}, 1125 (2023).

\bibitem{Close:1987er}
  F.~E.~Close,
  Gluonic Hadrons,
  Rept. Prog. Phys. \textbf{51}, 833 (1988).

\bibitem{Chen:2005mg}
  Y.~Chen, \textit{et al.},
  Glueball spectrum and matrix elements on anisotropic lattices,
  Phys. Rev. D \textbf{73}, 014516 (2006).

\bibitem{Athenodorou:2020ani}
  A.~Athenodorou and M.~Teper,
  The glueball spectrum of SU(3) gauge theory in 3+1 dimensions,
  J. High Energy Phys.  \textbf{11}, 172 (2020)

\bibitem{JPAC:2018zyd}
  A.~Rodas \textit{et al.} (JPAC Collaboration),
  Determination of the pole position of the lightest hybrid meson candidate,
  Phys. Rev. Lett. \textbf{122}, 042002 (2019).

\bibitem{Kopf:2020yoa}
  B.~Kopf \textit{et al.},
  Investigation of the lightest hybrid meson candidate with a coupled-channel analysis of ${{\bar{p}}p}$-, $\pi ^- p$- and ${\pi \pi }$-Data,
  Eur. Phys. J. C \textbf{81}, 1056 (2021).

\bibitem{BESIII:2022iwi}
  M.~Ablikim \textit{et al.} (BESIII Collaboration),
  Partial wave analysis of $J/\psi\rightarrow\gamma\eta\eta'$,
  Phys. Rev. D \textbf{106}, 072012 (2022).

\bibitem{BESIII:2022riz}
  M.~Ablikim \textit{et al.} (BESIII Collaboration),
  Observation of an isoscalar resonance with exotic $J^{PC}=1^{-+}$ quantum numbers in $J/\psi\rightarrow\gamma\eta\eta'$,
  Phys. Rev. Lett. \textbf{129}, 192002 (2022).

\bibitem{BESIII:2025izw}
  M.~Ablikim \textit{et al.} [BESIII],
  Search for $\eta_{1}(1855)$ in $\chi_{cJ}\to\eta\eta\eta^{\prime}$ decays,
  arXiv:2504.19087.

\bibitem{Farina:2020slb}
  C.~Farina, H.~Garcia Tecocoatzi, A.~Giachino, E.~Santopinto and E.~S.~Swanson,
  Heavy hybrid decays in a constituent gluon model,
  Phys. Rev. D \textbf{102}, 014023 (2020).

\bibitem{Swanson:2023zlm}
  E.~S.~Swanson,
  Light hybrid meson mixing and phenomenology,
  Phys. Rev. D \textbf{107}, 074028 (2023).

\bibitem{Woss:2020ayi}
  A.~J.~Woss \textit{et al.} (Hadron Spectrum Collaboration),
  Decays of an exotic $1^{-+}$ hybrid meson resonance in QCD,
  Phys. Rev. D \textbf{103}, 054502 (2021).

\bibitem{Liang:2024lon}
  J.~Liang, S.~Chen, Y.~Chen, C.~Shi and W.~Sun,
  Decay properties of light $1^{-+}$ hybrids from $N_f=2$ lattice QCD,
  Sci. China Phys. Mech. Astron. \textbf{68}, 251011 (2025).

\bibitem{Chen:2022isv}
  F.~Chen, X.~Jiang, Y.~Chen, M.~Gong, Z.~Liu, C.~Shi and W.~Sun,
  $1^{-+}$ hybrid meson in $J/\psi$ radiative decays from lattice QCD,
  Phys. Rev. D \textbf{107}, 054511 (2023).

\bibitem{Shi:2023sdy}
  C.~Shi, Y.~Chen, M.~Gong, X.~Jiang, Z.~Liu and W.~Sun,
  Decays of $1^{-+}$ charmoniumlike hybrid using lattice QCD,
  Phys. Rev. D \textbf{109}, 094513 (2024).

\bibitem{Schlosser:2025tca}
  C.~Schlosser and M.~Wagner,
  Hybrid spin-dependent and hybrid-quarkonium mixing potentials at order $(1/m_Q)^1$ from SU(3) lattice gauge theory,
  Phys. Rev. D \textbf{111}, 074504 (2025)

\bibitem{Wang:2023whb}
  Q.~N.~Wang, D.~K.~Lian and W.~Chen,
  Predictions of the hybrid mesons with exotic quantum numbers $J^{PC}=2^{+-}$,
  Phys. Rev. D \textbf{108}, 114010 (2023).

\bibitem{Tan:2024grd}
  W.~H.~Tan, N.~Su and H.~X.~Chen,
  Light single-gluon hybrid states with various exotic quantum numbers,
  Phys. Rev. D \textbf{110}, 034031 (2024).

\bibitem{Barsbay:2024vjt}
  B.~Barsbay, K.~Azizi and H.~Sundu,
  Light quarkonium hybrid mesons,
  Phys. Rev. D \textbf{109}, 094034 (2024).

\bibitem{Esmer:2025xss}
  G.~D.~Esmer, K.~Azizi, H.~Sundu and S.~T\"urkmen,
  Decays of the light hybrid meson $1^{-+}$,
  Phys. Rev. D \textbf{111}, 034041 (2025).

\bibitem{Wang:2024hvp}
  Z.~G.~Wang,
  Mass spectrum of the hidden-charm hybrid states via QCD sum rules,
  Phys. Rev. D \textbf{111}, 114009 (2025).

\bibitem{Barsbay:2025vjq}
  B.~Barsbay,
  Decays of the vector charmonium and bottomonium hybrids,
  arXiv:2506.00665.

\bibitem{Berwein:2024ztx}
  M.~Berwein, N.~Brambilla, A.~Mohapatra and A.~Vairo,
  Hybrids, tetraquarks, pentaquarks, doubly heavy baryons, and quarkonia in Born-Oppenheimer effective theory,
  Phys. Rev. D \textbf{110}, 094040 (2024).

\bibitem{Qiu:2022ktc}
  L.~Qiu and Q.~Zhao,
  Towards the establishment of the light $J^{P(C)}=1^{-(+)}$ hybrid nonet,
  Chin. Phys. C \textbf{46}, 051001 (2022).

\bibitem{Shastry:2022mhk}
  V.~Shastry, C.~S.~Fischer and F.~Giacosa,
  The phenomenology of the exotic hybrid nonet with $\pi_1(1600)$ and $\eta_1(1855)$,
  Phys. Lett. B \textbf{834}, 137478 (2022).

\bibitem{Chen:2022qpd}
  H.~X.~Chen, N.~Su and S.~L.~Zhu,
  QCD axial anomaly enhances the $\eta\eta^\prime$ decay of the hybrid candidate $\eta_{1}(1855)$,
  Chin. Phys. Lett. \textbf{39}, 051201 (2022).

\bibitem{Chen:2023ukh}
  B.~Chen, S.~Q.~Luo and X.~Liu,
  Constructing the $J^{P(C)}=1^{-(+)}$ light flavor hybrid nonet with the newly observed $\eta_1(1855)$,
  Phys. Rev. D \textbf{108}, 054034 (2023).

\bibitem{Dudek:2011bn}
  J.~J.~Dudek,
  The lightest hybrid meson supermultiplet in QCD,
  Phys. Rev. D \textbf{84}, 074023 (2011).

\bibitem{Dudek:2013yja}
  J.~J.~Dudek \textit{et al.} (Hadron Spectrum Collaboration),
  Toward the excited isoscalar meson spectrum from lattice QCD,
  Phys. Rev. D \textbf{88}, 094505 (2013).

\bibitem{Close:1994hc}
  F.~E.~Close and P.~R.~Page,
  The production and decay of hybrid mesons by flux tube breaking,
  Nucl. Phys. B \textbf{443}, 233 (1995).

\bibitem{Page:1998gz}
  P.~R.~Page, E.~S.~Swanson and A.~P.~Szczepaniak,
  Hybrid meson decay phenomenology,
  Phys. Rev. D \textbf{59}, 034016 (1999).

\bibitem{HadronSpectrum:2012gic}
  L.~Liu \textit{et al.} (Hadron Spectrum Collaboration),
  Excited and exotic charmonium spectroscopy from lattice QCD,
  J. High Energy Phys. \textbf{07}, 126 (2012).

\bibitem{Swanson:1998kx}
  E.~S.~Swanson and A.~P.~Szczepaniak,
  Heavy hybrids with constituent gluons,
  Phys. Rev. D \textbf{59}, 014035 (1999).

\bibitem{Godfrey:1985xj}
  S.~Godfrey and N.~Isgur,
  Mesons in a Relativized Quark Model with Chromodynamics,
  Phys. Rev. D \textbf{32}, 189 (1985).

\bibitem{Capstick:1986ter}
  S.~Capstick and N.~Isgur,
  Baryons in a relativized quark model with chromodynamics,
  Phys. Rev. D \textbf{34}, 2809 (1986).

\bibitem{Lucha:1991vn}
  W.~Lucha, F.~F.~Schoberl and D.~Gromes,
  Bound states of quarks,
  Phys. Rept. \textbf{200}, 127 (1991).

\bibitem{Mukherjee:1993hb}
 S.~N.~Mukherjee, \textit{et al.},
 Quark potential approach to baryons and mesons,
 Phys. Rept. \textbf{231}, 201 (1993).

\bibitem{Griffiths:1983ah}
  L.~A.~Griffiths, C.~Michael and P.~E.~L.~Rakow,
  Mesons With Excited Glue,
  Phys. Lett. B \textbf{129}, 351 (1983).

\bibitem{Campbell:1984fe}
  N.~A.~Campbell, L.~A.~Griffiths, C.~Michael and P.~E.~L.~Rakow,
  Mesons with excited glue from SU(3) lattice gauge theory,
  Phys. Lett. B \textbf{142}, 291 (1984).

\bibitem{Juge:1997nc}
  K.~J.~Juge, J.~Kuti and C.~J.~Morningstar,
  Gluon excitations of the static quark potential and the hybrid quarkonium spectrum,
  Nucl. Phys. B Proc. Suppl. \textbf{63}, 326 (1998).

\bibitem{Juge:2002br}
  K.~J.~Juge, J.~Kuti and C.~Morningstar,
  Fine structure of the QCD string spectrum,
  Phys. Rev. Lett. \textbf{90}, 161601 (2003)

\bibitem{Capitani:2018rox}
  S.~Capitani, O.~Philipsen, C.~Reisinger, C.~Riehl and M.~Wagner,
  Precision computation of hybrid static potentials in SU(3) lattice gauge theory,
  Phys. Rev. D \textbf{99}, 034502 (2019).

\bibitem{Schlosser:2021wnr}
  C.~Schlosser and M.~Wagner,
  Hybrid static potentials in SU(3) lattice gauge theory at small quark-antiquark separations,
  Phys. Rev. D \textbf{105}, 054503 (2022).

\bibitem{Alasiri:2024nue}
  F.~Alasiri, E.~Braaten and A.~Mohapatra,
  Born-Oppenheimer potentials for SU(3) gauge theory,
  Phys. Rev. D \textbf{110}, 054029 (2024).

\bibitem{Isgur:1984bm}
  N.~Isgur and J.~E.~Paton,
  Flux-tube model for hadrons in QCD,
  Phys. Rev. D \textbf{31}, 2910 (1985).

\bibitem{Kalashnikova:2002tg}
  Y.~S.~Kalashnikova and D.~S.~Kuzmenko,
  Hybrid adiabatic potentials in the QCD string model,
  Phys. Atom. Nucl. \textbf{66}, 955 (2003).

\bibitem{Barnes:1995hc}
  T.~Barnes, F.~E.~Close and E.~S.~Swanson,
  Hybrid and conventional mesons in the flux tube model: Numerical studies and their phenomenological implications,
  Phys. Rev. D \textbf{52}, 5242 (1995).

\bibitem{Jaffe:1975fd}
  R.~L.~Jaffe and K.~Johnson,
  Unconventional states of confined quarks and gluons,
  Phys. Lett. B \textbf{60}, 201 (1976).

\bibitem{Barnes:1982tx}
  T.~Barnes, F.~E.~Close and F.~de Viron,
  $Q\bar{Q}G$ hermaphrodite mesons in the MIT Bag model,
  Nucl. Phys. B \textbf{224}, 241 (1983)

\bibitem{ParticleDataGroup:2024cfk}
  S.~Navas \textit{et al.} (Particle Data Group),
  Review of particle physics,
  Phys. Rev. D \textbf{110}, 030001 (2024).

\bibitem{Ryan:2020iog}
  S.~M.~Ryan \textit{et al.} (Hadron Spectrum Collaboration),
  Excited and exotic bottomonium spectroscopy from lattice QCD,
  J. High Energy Phys. \textbf{02}, 214 (2021).

\bibitem{Karl:1999wq}
  G.~Karl and J.~E.~Paton,
  Gluelump spectrum in the bag model,
  Phys. Rev. D \textbf{60}, 034015 (1999).

\bibitem{Foster:1998wu}
  M.~Foster \textit{et al.} (UKQCD Collaboration),
  Hadrons with a heavy color-adjoint particle,
  Phys. Rev. D \textbf{59}, 094509 (1999).

\bibitem{Simonov:2000ky}
  Y.~A.~Simonov,
  Gluelump spectrum in the QCD string model,
  Nucl. Phys. B \textbf{592}, 350 (2001).

\bibitem{Guo:2007sm}
  P.~Guo, A.~P.~Szczepaniak, G.~Galata, A.~Vassallo and E.~Santopinto,
  Gluelump spectrum from Coulomb gauge QCD,
  Phys. Rev. D \textbf{77}, 056005 (2008).

\bibitem{Buisseret:2008pd}
  F.~Buisseret,
  Gluelump model with transverse constituent gluons,
  Eur. Phys. J. A \textbf{38}, 233 (2008).

\bibitem{Eichten:1978tg}
  E.~Eichten, K.~Gottfried, T.~Kinoshita, K.~D.~Lane and T.~M.~Yan,
  Charmonium: The Model,
  Phys. Rev. D \textbf{17}, 3090 (1978),
  Erratum: Phys. Rev. D \textbf{21}, 313 (1980).

\bibitem{Karl:1994ji}
  G.~Karl and V.~A.~Novikov,
  Variational estimates for excited states,
  Phys. Rev. D \textbf{51}, 5069 (1995).
  Erratum: Phys. Rev. D \textbf{55}, 4496 (1997).


\bibitem{Schoberl:1984ha}
  F.~Schoberl, P.~Falkensteiner and S.~Ono,
  Applications of a variational method to baryons and supersymmetric hadrons ($\tilde{t}Q\bar{Q}$ and $\tilde{g}Q\bar{Q}$),
  Phys. Rev. D \textbf{30}, 603 (1984).

\bibitem{Kalashnikova:1993xb}
  Y.~S.~Kalashnikova,
  Exotic hybrids and their non-exotic counterparts,
  Z. Phys. C \textbf{62}, 323 (1994).

\bibitem{Adams:2005tx}
  G.~S.~Adams \textit{et al.}  (E852 Collaboration),
  Exotic meson production in BNL experiment E852,
  J. Phys. Conf. Ser. \textbf{9}, 136 (2005).


\bibitem{E852:2004gpn}
  J.~Kuhn \textit{et al.} (E852 Collaboration),
  Exotic meson production in the $f_1(1285)\pi^-$ system observed in the reaction $\pi^-p\to\eta\pi^+\pi^-\pi^-p$ at 18 GeV$/c$,
  Phys. Lett. B \textbf{595}, 109 (2004).

\bibitem{E852:2004rfa}
  M.~Lu \textit{et al.} (E852 Collaboration),
  Exotic meson decay to $\omega\pi^0\pi^-$,
  Phys. Rev. Lett. \textbf{94}, 032002 (2005)

\bibitem{Meyer:2015eta}
  C.~A.~Meyer and E.~S.~Swanson,
  Hybrid Mesons,
  Prog. Part. Nucl. Phys. \textbf{82}, 21 (2015).

\bibitem{Swanson:1997wy}
  E.~S.~Swanson and A.~P.~Szczepaniak,
  Decays of hybrid mesons,
  Phys. Rev. D \textbf{56}, 5692 (1997).

\bibitem{Hayne:1981zy}
  C.~Hayne and N.~Isgur,
  Beyond the wave function at the origin: Some momentum-dependent effects in the nonrelativistic quark model,
  Phys. Rev. D \textbf{25}, 1944 (1982).

\bibitem{Chen:2016iyi}
  B.~Chen, K.~W.~Wei, X.~Liu and T.~Matsuki,
  Low-lying charmed and charmed-strange baryon states,
  Eur. Phys. J. C \textbf{77}, 154 (2017).

\bibitem{COMPASS:2014vkj}
  C.~Adolph \textit{et al.} (COMPASS Collaboration),
  Odd and even partial waves of $\eta\pi^-$ and $\eta'\pi^-$ in $\pi^-p\to\eta^{(\prime)}\pi^-p$ at $191\,\textrm{GeV}/c$,
  Phys. Lett. B \textbf{740}, 303 (2015),
  Erratum: Phys. Lett. B \textbf{811}, 135913 (2020).

\bibitem{Liu}
  Beijiang Liu, on behalf of the BESIII Colaboration,
  Light QCD exotics at BESIII,
  \href{https://indico.global/event/12671/contributions/112805/attachments/52078/100227/BJLiu_QCHSC2024.pdf}{XVIth Quark Confinement and the Hadron Spectrum, Queensland, Australia (2024)}.

%




\end{thebibliography}
\end{document}